\pdfoutput=1
\documentclass[9pt]{elife}

\usepackage{textcomp}
\microtypesetup{expansion=false}

\graphicspath{{figures/}}

\title{DRIADA: an open-source Python toolkit for linking single-neuron selectivity and population activity to behavior}

\author[1*]{Nikita Pospelov}
\author[1]{Viktor Plusnin}
\author[1]{Olga Rogozhnikova}
\author[1]{Anna Ivanova}
\author[2]{Vladimir Sotskov}
\author[1,3]{Margarita Orobets}
\author[1]{Ksenia Toropova}
\author[1]{Olga Ivashkina}
\author[4]{Vladik Avetisov}
\author[1]{Konstantin Anokhin}

\affil[1]{Institute for Advanced Brain Studies, Lomonosov Moscow State University, Moscow, Russia}
\affil[2]{Center for Interdisciplinary Research in Biology, Coll\`{e}ge de France, Paris, France}
\affil[3]{A.P. Nelyubin Institute of Pharmacy, I.M. Sechenov First Moscow State Medical University, Moscow, Russia}
\affil[4]{Semenov Institute of Chemical Physics, Russian Academy of Sciences, Moscow, Russia}

\corr{pospelov.na14@physics.msu.ru}{NP}

\begin{document}
\maketitle

\begin{abstract}
Understanding how individual neurons contribute to population activity during behavior is central to understanding brain functions across biological scales. We developed DRIADA, an open-source Python toolkit that connects analyses of individual neurons, population activity, and neuronal networks to the same behavioral parameters. We evaluated it on synthetic populations with known ground truth, calcium-imaging recordings from dorsal hippocampal CA1 in 13~mice across 39~sessions, and a simulated continuous attractor neural network (CANN) with known toroidal organization. Across most hippocampal sessions, embeddings of spatially selective neurons predicted held-out position better than equally sized random subsets and all recorded neurons. The same workflow related population dimensions back to individual neurons, including cells without detected selectivity for the measured behavioral variables. In the simulated network, complementary analyses captured features consistent with its known organization. Thus, DRIADA provides a common framework for relating neuronal selectivity, population structure, and network activity to behavioral dynamics.
\end{abstract}

\section{Introduction}

Single-neuron and population-level approaches provide complementary views of neural activity, but are rarely integrated within a common analytical framework. The tuning curves and selectivity
profiles characterize how individual cells vary their activity in relation to experimentally defined sensory, spatial, or behavioral variables  \citep{hubel1962receptive, okeefe1971hippocampus}.  At the population level, complementary approaches capture different aspects of collective organization: dimensionality
reduction identifies the low-dimensional manifolds along which
activity evolves \citep{cunningham2014dimensionality, chung2021neural},  while graph-theoretic analysis describes modular functional organization and
its associations with behavior
\citep{bullmore2009complex, sporns2005human}.  In practice, however,
each paradigm is typically pursued with its own tools, data formats, and conventions, so their integration requires combining unrelated codebases and translating between conceptual frameworks
\citep{vaccari2022single, saxena2019towards}.

Although these levels of analysis are conceptually distinct, relating them is important for understanding neural coding. Some studies have linked them, for example by connecting single-neuron tuning
to manifold structure \citep{chung2021neural}, but cross-level studies remain rare, and the tooling gap mirrors the
conceptual gap.  Recent work suggests that this fragmentation can change scientific conclusions.
\citet{munn2024scaling} showed that the distinct coding signatures observed at different recording resolutions emerge from a common multi-scale organization of neural activity, indicating that scale-dependent theories of brain function describe different facets of a single underlying structure rather than distinct phenomena. \citet{vaccari2022single} argued that population-level analyses
remain incomplete without characterizing the individual neurons that
compose the population. \citet{dubreuil2022role} using low-rank recurrent neural network (RNN) models demonstrated that the single-neuron subpopulation structure and the dimensionality of population dynamics play complementary roles in shaping network computations. \citet{esparza2025celltype} reported a parallel finding in the mouse hippocampus: cell-type specific subpopulations of CA1 neurons can exhibit distinct manifold geometries that are obscured when the entire hippocampal population is analyzed jointly.  These findings raise questions that span levels of analysis: Is single-neuron selectivity reflected in network membership? Does filtering by
selectivity improve manifold reconstruction? Do all neurons identified as selective by single-neuron analysis contribute equally to population structure? Yet answering them with current tools demands custom pipelines that complicate analysis.

The need for integration is amplified by the scale and nature of modern recordings. Recordings now scale to populations of tens of thousands to a million neurons \citep{stringer2019high, manley2024simultaneous}, while complex behavioral tasks rely on richer population codes \citep{rigotti2013importance, fusi2016why, tye2024mixed}. Calcium imaging adds further constraints: it produces continuous fluorescence signals shaped by
slow indicator kinetics \citep{climer2021information}. These trends make a unified
software workflow a practical prerequisite for cross-scale
investigation.

Several packages address subsets of this analysis space, each
operating within a single paradigm (Table~\ref{tab:comparison}).

\textbf{Information-theoretic toolboxes.}
NIT \citep{maffulli2022nit} provides mutual information (MI) estimation,
including Gaussian-copula mutual information (GCMI) for continuous signals and extensive bias corrections,
but lacks autocorrelation-aware permutation testing, automatic
variable type detection, or integration with dimensionality reduction
and network analysis.  MINT (Multivariate Information in Neuroscience Toolbox) \citep{lorenz2025mint} combines MI, partial information decomposition (PID),
and feature-specific information transfer with dimensionality
reduction in a MATLAB implementation.  FRITES (Framework for Information
Theoretical analysis of Electrophysiological data and Statistics)
\citep{combrisson2022frites} offers GCMI-based MI workflows with
group-level non-parametric permutation statistics for
neurophysiological data.  Information Dynamics Toolkit~xl (IDTxl) \citep{wollstadt2019idtxl} and HOI
\citep{neri2024hoi} focus on information dynamics and higher-order
interactions, respectively.  To our knowledge, no existing package integrates single-neuron
information testing, population manifold structure, and network analysis under a single Python data model.

\textbf{Spike-train and connectivity tools.}
Elephant \citep{denker2024elephant} (Electrophysiology Analysis Toolkit)
provides spike-train statistics and connectivity estimation, but does not
address population manifolds, single-neuron selectivity testing, or
time-series--derived networks.

\textbf{Population-level tools.}
CEBRA (Consistent EmBeddings of high-dimensional Recordings using Auxiliary variables) \citep{schneider2023learnable} produces consistent embeddings through contrastive learning and operates at the population level. CILDS (Calcium Imaging Linear Dynamical
System)
\citep{koh2023dimensionality} jointly performs deconvolution and reduces
dimensionality without selectivity analysis. Demixed~PCA
\citep{kobak2016demixed} decomposes population activity by categorical task variable and
provides encoder weights to quantify each neuron's contribution to each component; its significance testing applies at the level of demixed components (via shuffling-based classification) rather than to individual neurons, and its scope is restricted to categorical task variables.

\textbf{Data frameworks.}
Pynapple \citep{viejo2023pynapple} provides a typed time-series
framework for neurophysiology with tuning curve computation and
decoding, with scope limited to data manipulation and basic analyses. CaImAn
\citep{giovannucci2019caiman} handles upstream calcium imaging
preprocessing and is complementary to DRIADA\footnote{\url{https://github.com/iabs-neuro/driada}}, which begins where
preprocessing ends.

\begin{table}[bt]
\caption{\label{tab:comparison}Feature comparison of DRIADA with related
neural data analysis tools.  Tools selected for inclusion provide at
least one of: single-neuron MI testing, population dimensionality
reduction, or functional network analysis.  NIT
\citep{maffulli2022nit} provides MI estimation; MINT
\citep{lorenz2025mint} analyzes population-level information
transmission; FRITES \citep{combrisson2022frites} provides
MI-based neurophysiological statistics; CEBRA
\citep{schneider2023learnable} learns population embeddings; dPCA
\citep{kobak2016demixed} decomposes activity by task variable;
Elephant \citep{denker2024elephant} provides spike train statistics
and connectivity estimation.}
\begin{tabular}{l c c c c c c c}
\toprule
Feature & DRIADA & NIT & MINT & FRITES & CEBRA & dPCA & Elephant \\
\midrule
Role of individual neurons      & $\checkmark$ & $\checkmark$ & $\checkmark$          & ---          & ---          & $\checkmark$ & ---          \\
Autocorrelation-aware p-values  & $\checkmark$ & ---          & ---          & $\checkmark$          & ---          & ---          & ---          \\
Diverse variable types          & $\checkmark$ & $\checkmark$ & ---          & $\checkmark$          & $\checkmark$ & ---          & ---          \\
Population coding analysis      & $\checkmark$ & ---          & $\checkmark$ & $\checkmark$ & $\checkmark$ & $\checkmark$ & $\checkmark$ \\
Network analysis                & $\checkmark$ & ---          & ---          & $\checkmark$ & ---          & ---          & $\checkmark$ \\
Network analysis of time series & $\checkmark$ & ---          & ---          & ---          & ---          & ---          & ---          \\
\bottomrule
\end{tabular}
\end{table}

An additional analytical dimension comes from complex systems science:
constructing networks directly from individual time series
\citep{varley2022network}.  Rather than preserving spatial information
and collapsing time (as in functional connectivity), these methods
preserve temporal structure, mapping a single neuron's state-space
trajectory into a graph.  Recurrence networks capture how a signal revisits in phase-space using
delay embedding
\citep{donner2010recurrence, marwan2007recurrence}; at the population
level, pairwise similarity between per-neuron recurrence graphs yields
functional networks based on temporal structure rather than
activity covariance. Visibility graphs encode extreme-event structure
without free parameters \citep{lacasa2008time, luque2009horizontal},
and ordinal partition networks quantify temporal complexity through
symbolic dynamics \citep{bandt2002permutation, mccullough2015time}.
All three have been applied to neural data at the
mesoscale (\citet{varley2022network} demonstrated all three on
electrocorticography, ECoG, recordings), but applications at single-neuron resolution
remain limited, with most reported uses in neuroscience concentrated at the electroencephalography / local field potential (EEG/LFP) scale \citep{sulaimany2023visibility}. To our knowledge, no integrated package uses all three method families within a neuroscience-oriented workflow. Existing implementations either cover a single
method (e.g., \texttt{ts2vg}\footnote{\url{https://github.com/CarlosBergillos/ts2vg}}
for visibility graphs) or span multiple methods but are developed primarily for climate and complex-systems applications  (e.g.,
\texttt{pyunicorn}; \citealp{donges2015unified}).

Each of these tools covers one part of the analysis; what is missing are the connections between them. To our knowledge, no Python package integrates neuronal selectivity testing, manifold extraction, and network analysis within a single data model.  Beyond this integration gap,  the Python ecosystem for
graph-analytic brain network analysis is also narrow: the Brain
Connectivity Toolbox \citep{rubinov2010bct} remains the reference
implementation for graph metrics in neuroscience, but is MATLAB-based
and operates on pre-constructed connectivity matrices, without tools for
building networks from time series (visibility, recurrence, or
ordinal-partition graphs) or for integrating single-neuron and
population-level analyses; its Python
ports\footnote{\url{https://github.com/aestrivex/bctpy},
\url{https://github.com/fiuneuro/brainconn}} have received little
development in recent years.

DRIADA fills this gap with a shared data model that makes all
analysis modules interoperable.  Six modules operate on a common
hierarchy of typed data objects:
(1)~single-neuron selectivity testing via GCMI
\citep{ince2017statistical} and Kraskov--St\"{o}gbauer--Grassberger (KSG) \citep{kraskov2004estimating}
estimators with autocorrelation-aware circular-shift permutation
testing;
(2)~dimensionality reduction that spans linear, spectral, and
autoencoder approaches (15 methods);
(3)~network analysis (functional connectivity, spectral
decomposition, community detection, and thermodynamic entropy);
(4)~representational similarity analysis
\citep{kriegeskorte2008representational};
(5)~information-theoretic utilities for custom MI and entropy
computation;
(6)~network analysis of time series (recurrence networks, visibility
graphs, ordinal partition networks) \citep{varley2022network}.
DRIADA implements all three time-series network methods alongside
selectivity testing, dimensionality reduction, and network analysis;
this paper demonstrates recurrence networks in a cross-scale workflow
(Figure~\ref{fig:cann}D).

Behavioral and environmental
variables (sensory inputs, motor outputs, and internal states relevant to neural activity) are
stored in the same data model as typed
\emph{TimeSeries}, alongside the neural signals.  As a result, selectivity testing, embedding analysis, and network construction all
have native access to behavior.
Variables undergo automatic type detection (continuous, discrete,
circular, multivariate), so that appropriate estimators and
preprocessing are selected automatically.  Selectivity
results can filter population matrices before dimensionality
reduction; embedding coordinates become features for reverse
selectivity analysis; graph-based dimensionality reduction returns
proximity graphs that inherit the full network toolkit; multiple
latent representations can be computed, compared, and linked back to
individual neurons within a single experiment; and the same
MI-based significance core underlies cell--variable, cell--cell, and
variable--variable testing. Although this paper focuses on calcium
imaging, the underlying data model is substrate-agnostic, processing
calcium traces, spike trains, and artificial neural network
activations through the same typed time-series abstraction.  We wrote DRIADA as a new package because integrating selectivity, population, and network analyses requires a shared underlying data model, which is not achievable through plugins to frameworks built around a single analytical layer.

We evaluated DRIADA through four increasingly complex demonstrations.
We first validated the full pipeline on a synthetic population with known
ground truth, then applied it to the hippocampal calcium imaging of
13~mice to test whether it recovers established selectivity profiles and to characterize
the broader neuronal selectivity profile of CA1. We then illustrated the
cross-scale transitions that motivate the shared data model: using
selectivity to reshape manifold geometry and using embedding
coordinates to interrogate single-neuron roles.  Finally, we simulated a toroidal continuous attractor network and
asked whether four independent modules (selectivity testing,
autoencoder embedding, representational similarity, and recurrence
networks) converge on the same nonlinear topology.

\section{Results}

Unless stated otherwise, values are reported as mean $\pm$ half-width of the 95\% confidence interval across sessions.

\begin{figure}[t!]
\begin{fullwidth}
\includegraphics[width=0.95\linewidth]{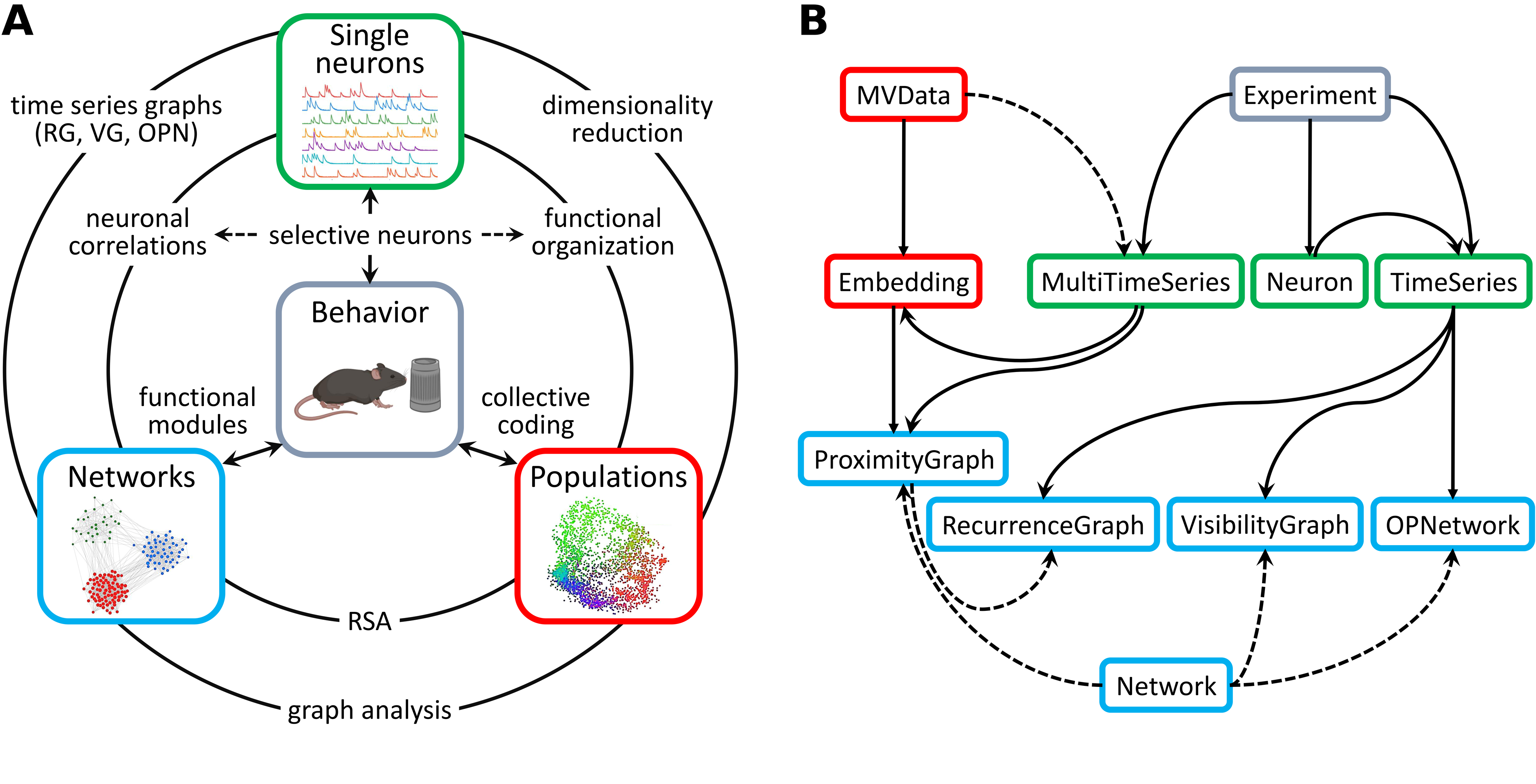}
\caption{\textbf{DRIADA architecture and data model.}
\textbf{(A)}~Analysis workflow. Complementary
perspectives, single-neuron characterization (green), population-level
geometry (red), and network structure (blue), form a triangle of
cross-scale connections mediated by shared behavioral data (center).
Outer arcs show analysis pathways that bridge scales:
time series graphs (recurrence, visibility, ordinal partition)
link single neurons to networks;
dimensionality reduction links single neurons to populations;
graph analysis links populations to networks.
Inner arcs show complementary routes: neuronal correlations (pairwise MI
between neurons) link single neurons to networks; functional organization,
the reverse analysis of each neuron's contribution to population dimensions,
links single neurons to populations; representational similarity analysis
(RSA) links populations to networks.
Arrows at the center show behavioral anchoring:
selectivity testing identifies individual neurons encoding specific
behavioral variables, and these neurons enter the correlation and
functional-organization analyses (dashed arrows); collective coding and functional modules
connect behavior to population and network representations
respectively.
\textbf{(B)}~Software object relationships.  Each analysis
perspective maps onto a set of interoperable Python classes.
The \emph{Experiment} container (gray) holds neural data as
\emph{MultiTimeSeries} and behavioral variables as
\emph{TimeSeries}; each \emph{Neuron} stores its own traces
(green).
\emph{MVData} provides dimensionality reduction, producing an
\emph{Embedding} that contains a \emph{ProximityGraph} (red).
Single-neuron time series generate recurrence graphs,
visibility graphs, and ordinal partition networks, all inheriting
the full \emph{Network} analysis toolkit (blue).
Solid arrows indicate \emph{produces} or \emph{contains}
relationships; dotted arrows indicate class inheritance.}
\label{fig:architecture}
\end{fullwidth}
\end{figure}

\subsection{A unified data model for cross-scale neural analysis}
\label{sec:results-architecture}

DRIADA addresses the fragmented-toolchain problem described in the Introduction through a shared data model that makes all analysis modules interoperable
(Figure~\ref{fig:architecture}).  The six
analysis modules are organized around complementary
perspectives (single-neuron characterization, population-level
geometry, and network structure) that form a triangle of cross-scale
connections mediated by shared behavioral data
(Figure~\ref{fig:architecture}A).

The \emph{Experiment} object is the central container: it holds
the neural activity matrix (raw calcium fluorescence or spike counts),
time-aligned behavioral variables, and per-neuron metadata in a
single structure from which any module can draw
(Figure~\ref{fig:architecture}B).
Each neuron's activity is wrapped as a \emph{TimeSeries}; the full
population is jointly accessible as a \emph{MultiTimeSeries} that
inherits dimensionality-reduction capabilities from its parent class
\emph{MVData}.
Behavioral variables undergo automatic type detection upon loading:
continuous features (e.g., position, speed), discrete features (e.g.,
object interaction, behavioral state), circular features (e.g., head
direction), and multivariate features (e.g., two-dimensional
coordinates) are classified and preprocessed appropriately: copula
normalization for GCMI, integer coding for discrete entropy, and
$(\cos,\sin)$ encoding for circular variables.  This automatic
handling avoids a common source of errors: applying an
inappropriate MI estimator to a misclassified variable.

\textbf{Single-neuron analysis} characterizes individual neurons from
two complementary angles.  Selectivity testing via INTENSE (Information-theoretic
Neuron--feature Selectivity Testing; \citealp{Pospelov2026intense},
integrated into DRIADA) identifies \emph{what} behavioral features each neuron's activity covaries with, that is, which variables it carries
significant information about.  Network analysis of time series
characterizes \emph{how} each neuron's dynamics unfold:
recurrence networks reveal dynamical macro-states via delay-embedded
phase space topology; visibility graphs capture extreme-event
structure and dynamical regime without free parameters; and ordinal
partition networks quantify temporal complexity via permutation
entropy.  All three graph types
inherit the full network analysis toolkit, so spectral
decomposition, community detection, and entropy measures are
immediately available on any single-neuron trace.

\textbf{Population activity analysis} compresses high-dimensional
neural recordings into interpretable low-dimensional representations.
Dimensionality reduction provides 15 methods spanning linear (PCA),
spectral graph-based (Laplacian eigenmaps, diffusion maps, Isomap),
neighborhood-based (t-SNE, UMAP), and neural network-based
(autoencoder, variational autoencoder) approaches, plus intrinsic
dimensionality estimators.  Representational similarity analysis (RSA)
compares neural representations across conditions via dissimilarity
matrices, with bootstrap estimates of uncertainty.

\textbf{Network analysis} characterizes collective organization
through functional connectivity, spectral decomposition, community
detection, and thermodynamic entropy measures.  The \emph{Network}
class provides a general-purpose graph-analysis framework that accepts
sparse adjacency matrices or NetworkX graphs.

These three perspectives are connected by six data-flow pathways
(Figure~\ref{fig:architecture}A, arcs).  From single neurons to
populations, individual traces compose into a population matrix that
feeds directly into any of the dimensionality reduction methods;
selectivity results from INTENSE can filter the population before
embedding, allowing direct comparison of full-population
versus selective-population manifolds.  Conversely, reverse selectivity analysis
tests single neurons against each embedding dimension, giving each neuron's
contribution to the population manifold.  From single neurons to
networks, pairwise MI significance testing between neurons yields
functional connectivity graphs, while per-neuron recurrence graphs,
visibility graphs, and ordinal partition networks transform
individual traces into graph representations whose pairwise
similarity produces neuron--neuron networks that can be analyzed with the same community detection toolkit. From populations to networks, graph-based dimensionality
reduction methods return a proximity graph that inherits the full
network toolkit (spectral analysis, inverse participation ratio, entropy), so population
manifold structure can be analyzed as a network;
representational dissimilarity matrices provide a complementary
route to the same network tools.  A shared information-theoretic
core (MI, conditional MI, entropy, time-delayed MI) underlies both the
selectivity-testing and recurrence modules, providing a uniform
statistical framework across scales.

\subsection{Recovering known structure from synthetic populations}
\label{sec:results-synthetic}

To validate DRIADA's end-to-end pipeline, we generated a synthetic
population with known ground-truth selectivity and modular network
structure (see Methods).  The population comprised two subpopulations:
(i)~300 event-modulated neurons responding to three discrete
behavioral events (E1, E2, E3): 180 single-event neurons
(60 per event), 60 mixed-selectivity neurons responding to pairs of
events (20 each for E1$\cup$E2, E1$\cup$E3, E2$\cup$E3), and 60
non-selective neurons; and (ii)~60 head-direction (HD) neurons with
von~Mises tuning, along with 40 additional non-selective
neurons.  The full population comprised 400 neurons (simulation parameters in Methods).

\begin{figure}
\begin{fullwidth}
\includegraphics[width=0.95\linewidth]{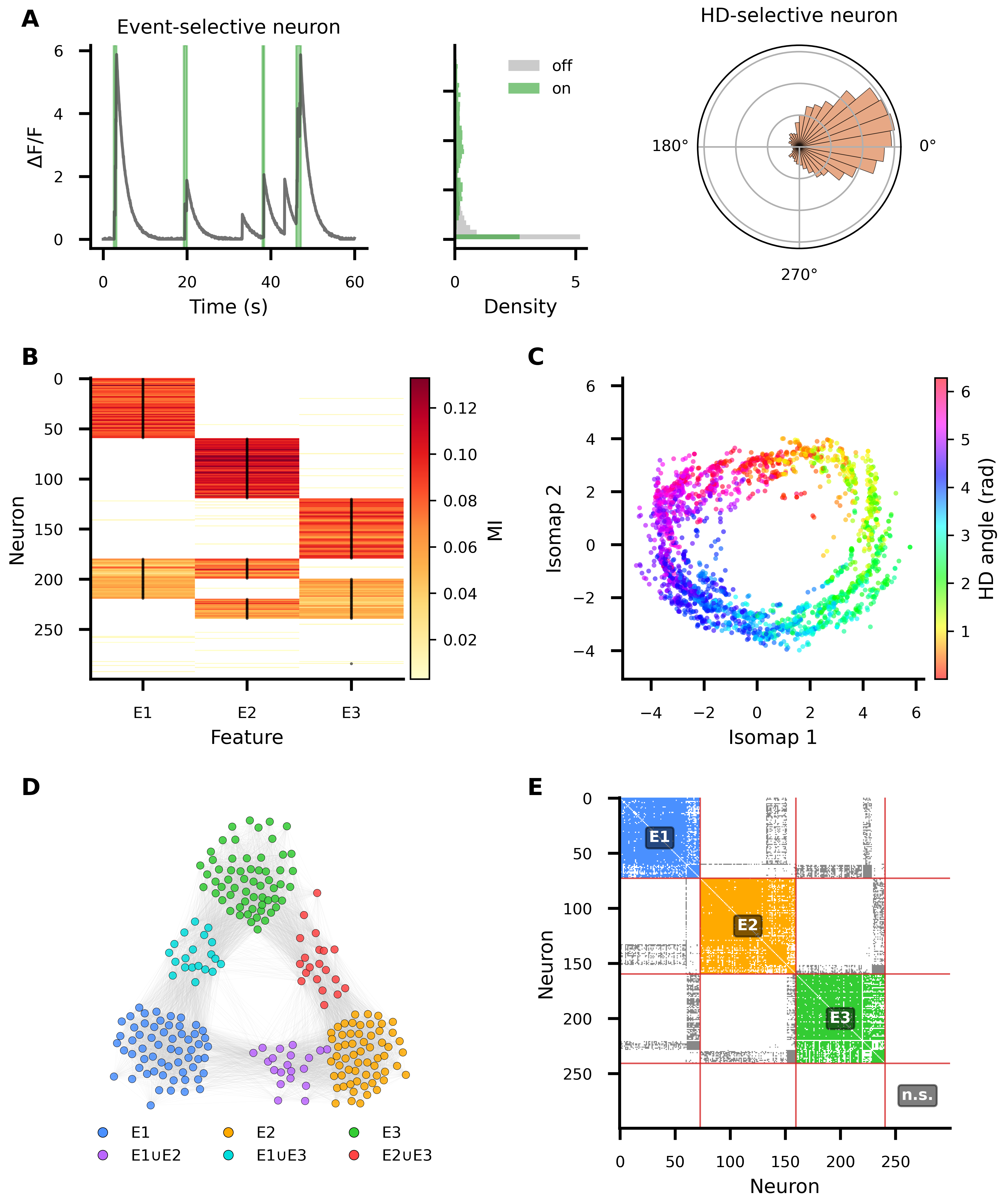}
\caption{\textbf{End-to-end validation on synthetic data with known
ground truth.}
\textbf{(A)}~Example selective neurons: an event-selective neuron
(left) with calcium trace during event epochs (green) and
corresponding density comparison (middle); a head-direction-selective
neuron shown as a polar tuning curve (right).
\textbf{(B)}~MI selectivity heatmap for 300 event neurons across
three discrete features (E1, E2, E3); significant pairs marked.
\textbf{(C)}~Isomap embedding of 60 HD neurons colored by head
direction angle, recovering the circular manifold.
\textbf{(D)}~Spring-layout network graph of event neurons colored
by six ground-truth groups; mixed-selectivity neurons bridge
pure-selectivity clusters.
\textbf{(E)}~Community-sorted adjacency matrix; within-community
connections colored by the dominant ground-truth group of the community, cross-community connections
in gray.}
\label{fig:synthetic}
\end{fullwidth}
\end{figure}

\textbf{Example selective neurons} (Figure~\ref{fig:synthetic}A).
Two representative neurons illustrate distinct selectivity types
recovered by the pipeline.  An event-selective neuron (left) shows
elevated calcium transients during event epochs (green shading), with
the corresponding density comparison (middle) confirming higher
activity during on- versus off-periods.  A head-direction-selective
neuron (right) exhibits clear angular tuning in a polar plot of mean
calcium activity as a function of head direction.

\textbf{Selectivity detection} (Figure~\ref{fig:synthetic}B).
The INTENSE pipeline was applied to the 300 event-modulated neurons
across three discrete features.  The MI heatmap shows mutual
information values for all neuron--feature pairs, with significant
pairs marked (black dots; two-stage test,
$p < 0.01$, Holm correction; see Methods).  INTENSE detected all 300 planted
pairs with one false positive (F1 $>$ 0.99), recovering the ground-truth modular selectivity
structure including mixed-selectivity neurons.

\textbf{Manifold recovery} (Figure~\ref{fig:synthetic}C).
Isomap dimensionality reduction of the 60 HD-neuron subpopulation
recovered the expected circular manifold.  The two-dimensional
embedding colored by ground-truth head direction angle shows a
continuous ring topology, confirming that the population geometry
preserves the circular behavioral variable.
The circular--circular correlation between embedding angle and
ground-truth head direction was $\rho_{\mathrm{circ}} = 0.90$
(circular-shift permutation test,
$p = 0.001$; see Methods).

\textbf{Functional network and community detection}
(Figure~\ref{fig:synthetic}D--E).
Cell--cell MI significance testing on the event-modulated
subpopulation yielded a functional connectivity network with
about 10{,}000 significant edges among 300 neurons
(edge density 0.23).  The giant connected component comprised
241 of 300 neurons; the 59 remaining isolates belong to the
non-selective subpopulation (60 planted units), which received almost no
significant inter-neuron connections.  The spring-layout
graph (panel~D) is colored by six ground-truth neuron types:
three pure-selectivity groups (E1, E2, E3) and three
mixed-selectivity groups (E1$\cup$E2, E1$\cup$E3, E2$\cup$E3).
Mixed-selectivity neurons lie between the
pure-selectivity clusters they bridge.
Louvain community detection (resolution $\gamma = 1.0$, fixed random seed) recovered three communities
corresponding to the three event types, plus 59 unclustered
neurons closely matching the 60 planted non-selective units
(adjusted Rand index, ARI = 0.72; normalized mutual information,
NMI = 0.78).
The community-sorted adjacency matrix (panel~E) shows within-community
connections colored by the dominant ground-truth group of their community, with
cross-community connections in gray.  Block-diagonal structure
shows that the detected communities align with the planted modular
organization; off-diagonal connections near block boundaries
correspond to mixed-selectivity neurons.

\subsection{Hippocampal selectivity profile}
\label{sec:results-realdata}

After validating  the pipeline on synthetic data with known ground
truth, we next asked whether DRIADA could recover established
biological findings from real hippocampal calcium imaging recordings
and characterize the broader selectivity profile of CA1.  We applied the
full pipeline to a dataset of 13 mice with one-photon miniscope
recordings from dorsal CA1 during free exploration of a circular
open-field arena (see Methods).
For each session, we extracted 11 behavioral features, spanning
continuous variables (two-dimensional position, head
direction, body direction, speed) and discrete behavioral states
(run, walk, rest, freezing, rearing, walls, center).  We illustrate the pipeline on an example session,
the one with the largest number of recorded neurons (695; range 26--695 across sessions),
and report cross-animal statistics from 39 sessions across all
13 mice.

\begin{figure}[p]
\begin{fullwidth}
\includegraphics[width=0.95\linewidth]{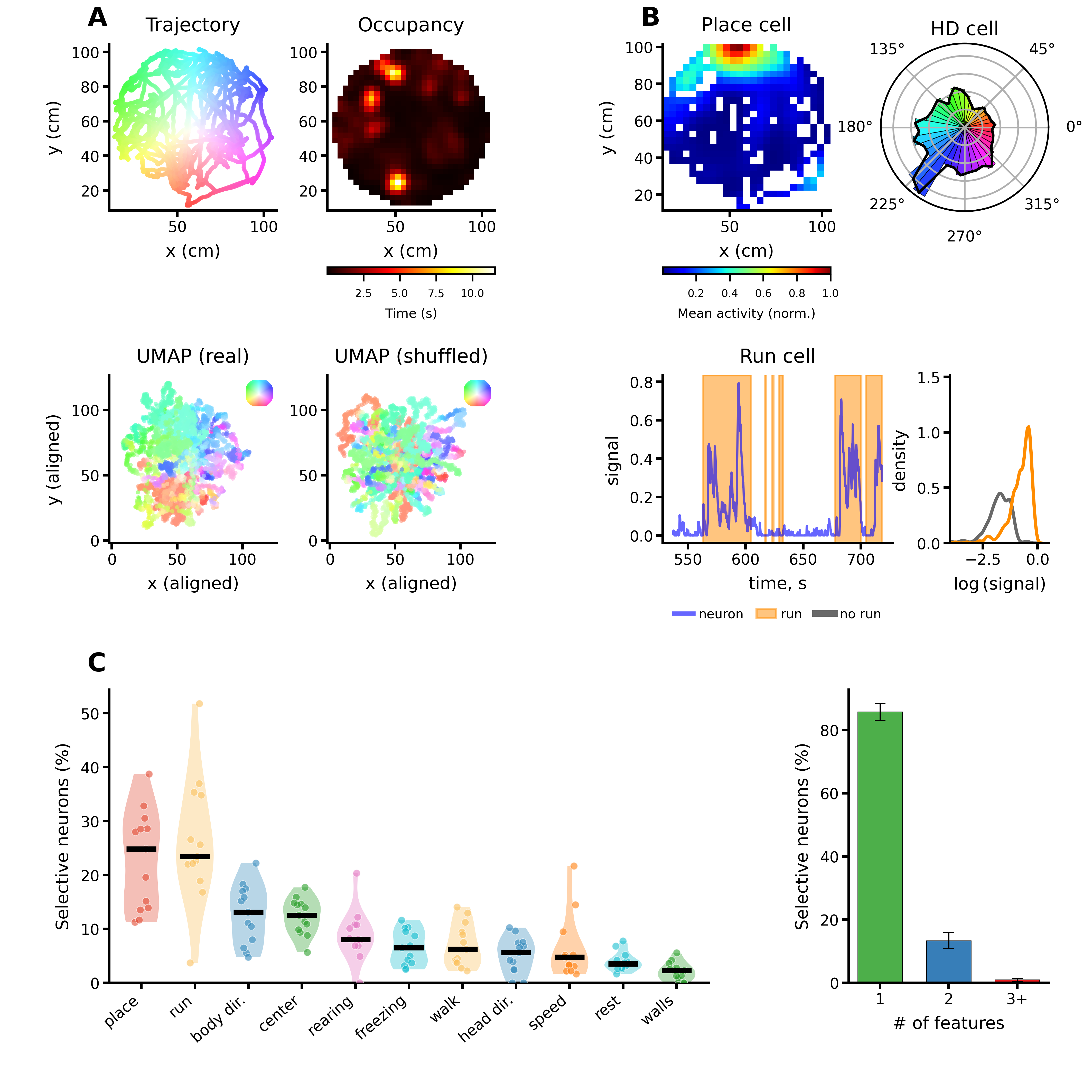}
\caption{\textbf{DRIADA pipeline applied to hippocampal calcium
imaging recordings during open-field exploration.}
\textbf{(A)}~Dataset overview for the example session: animal
trajectory colored by two-dimensional position (top left), spatial
occupancy heatmap (top right), UMAP embedding of spatially selective
neurons Procrustes-aligned to physical coordinates (bottom left), and
the same embedding computed from temporally shuffled data (bottom
right).
\textbf{(B)}~Example selective neurons: spatial calcium activity map
of a place cell (top left), polar tuning curve of a head-direction
cell (top right), and calcium trace overlaid with runs
for a run-modulated cell (bottom left) with the corresponding
activity-density comparison (bottom right).
\textbf{(C)}~Selectivity prevalence across 13~mice (left): violin plots show, among selective neurons, the percentage selective to each of 11~behavioral features, with individual mice as data points and
median indicated by a horizontal bar; features ordered by median.  Bar chart (right): percentage
of selective neurons selective to one, two, or three or more features
(mean $\pm$ 95\% CI across sessions).
}
\label{fig:realdata}
\end{fullwidth}
\end{figure}

\textbf{Dataset overview and population embedding}
(Figure~\ref{fig:realdata}A).  The animal's trajectory visited 79\% of the arena (3.5-cm bins).  To visualize population-level spatial structure, we applied UMAP to the calcium activity matrix of
neurons identified by INTENSE as carrying significant information about position,
wall-zone, or center-zone occupancy, and aligned the resulting
two-dimensional embedding to the animal's physical trajectory via
Procrustes analysis (see Methods).  The embedding preserves the
spatial layout of the arena (Procrustes disparity $= 0.41$, where $0$ indicates perfect
correspondence and $1$ no correspondence): points colored by two-dimensional position
reproduce the layout of the arena.  A control
embedding computed from temporally shuffled calcium traces
(independent circular shifts per neuron; see Methods) shows no
spatial organization (disparity $= 0.97$), indicating that the observed
manifold structure depends on intact temporal relationships in neural activity.

\textbf{Example selective neurons}
(Figure~\ref{fig:realdata}B).  Three representative neurons
illustrate distinct selectivity types detected by INTENSE: a place
cell with a spatially localized activity map; a head-direction cell
with a clear angular tuning curve shown as a polar plot; and a
run-modulated cell whose calcium trace is elevated during runs (orange shading), with the accompanying density plot confirming higher activity during runs than outside them.

\textbf{Selectivity prevalence across the population}
(Figure~\ref{fig:realdata}C).  To characterize the full range of
selectivity, we applied INTENSE to all 11~behavioral features across
all 39~sessions and summarized the results using DRIADA's
cross-analysis database.  For each mouse (averaging across sessions),
we computed the percentage of selective neurons that were significantly selective to each
feature (Figure~\ref{fig:realdata}C, left).  Selectivity was widespread: eight features (run, place, body direction,
center-zone occupancy, walk, freezing, speed, and rest) had at least one significantly selective neuron in
every animal, and the remaining three had it in 11--12 of 13 mice (per-pair threshold $p < 0.001$; see \citealp{Pospelov2026intense} for
threshold selection). Overall, CA1 selectivity extended beyond the classically emphasized place and head-direction neurons to most of the measured behavioral variables.
Fast locomotion (run) and place were the two most prevalent
selectivities (26~$\pm$~7\% and 23~$\pm$~6\% of selective neurons across
mice), followed by body direction
(13~$\pm$~3\%) and center-zone occupancy (12~$\pm$~2\%).
Head-direction selectivity (5~$\pm$~2\%) was lower than
typical proportions in head-direction-specialized regions \citep{taube1990head, taube1995head}, consistent with the weaker HD modulation reported
in dorsal CA1 \citep{leutgeb2000convergence, acharya2016causal}; part of
this difference follows from the disentanglement stage, which removes the
head-direction label from neurons also selective to body direction
(253 head-direction labels before the stage, 171 after).  Wall-zone selectivity was least prevalent (2~$\pm$~1\%). The rank ordering of feature prevalences was significantly
concordant across all 13 mice (Kendall's $W = 0.63$,
$p < 0.001$).
Selectivity labels of the same neurons were rarely retained across all
three daily sessions, but more often than expected by chance (see Methods).

The accompanying bar chart (Figure~\ref{fig:realdata}C, right)
shows the distribution of multi-feature selectivity. Of the selective neurons, 86~$\pm$~3\% were selective exactly to
one feature, 13~$\pm$~3\% to two features, and 1.0~$\pm$~0.5\% to three or more.
These counts were taken after the disentanglement stage of INTENSE
(see Methods), which removes associations explained by covarying
behavioral variables.  The $14\%$
of selective neurons that remain selective to two or more features after
disentanglement are mixed-selective with respect to the measured variables
\citep{rigotti2013importance, fusi2016why}.

Thus, the pipeline recovered known properties of hippocampal coding.  Overall, 24\% of neuron-sessions were classified as selective to at least one feature at $p < 0.001$.  The scale-bridging analysis
below asks whether this selective minority accounts for
population-level spatial structure, or whether the remaining
nominally non-selective neurons also contribute.

\subsection{Scale-bridging: from selectivity to population structure and functional networks}
\label{sec:results-scalebridging}

A central design goal of DRIADA is to connect single-neuron
selectivity with population-level and network-level structure.  The
shared data model makes these connections explicit: selectivity
results can filter the population before dimensionality reduction,
embedding coordinates can be treated as features for reverse
selectivity analysis, and pairwise MI testing reveals functional
connectivity.  We demonstrate these cross-scale transitions on the
same hippocampal dataset described above.

\begin{figure}
\begin{fullwidth}
\includegraphics[width=0.85\linewidth,trim=0 8bp 0 7bp,clip]{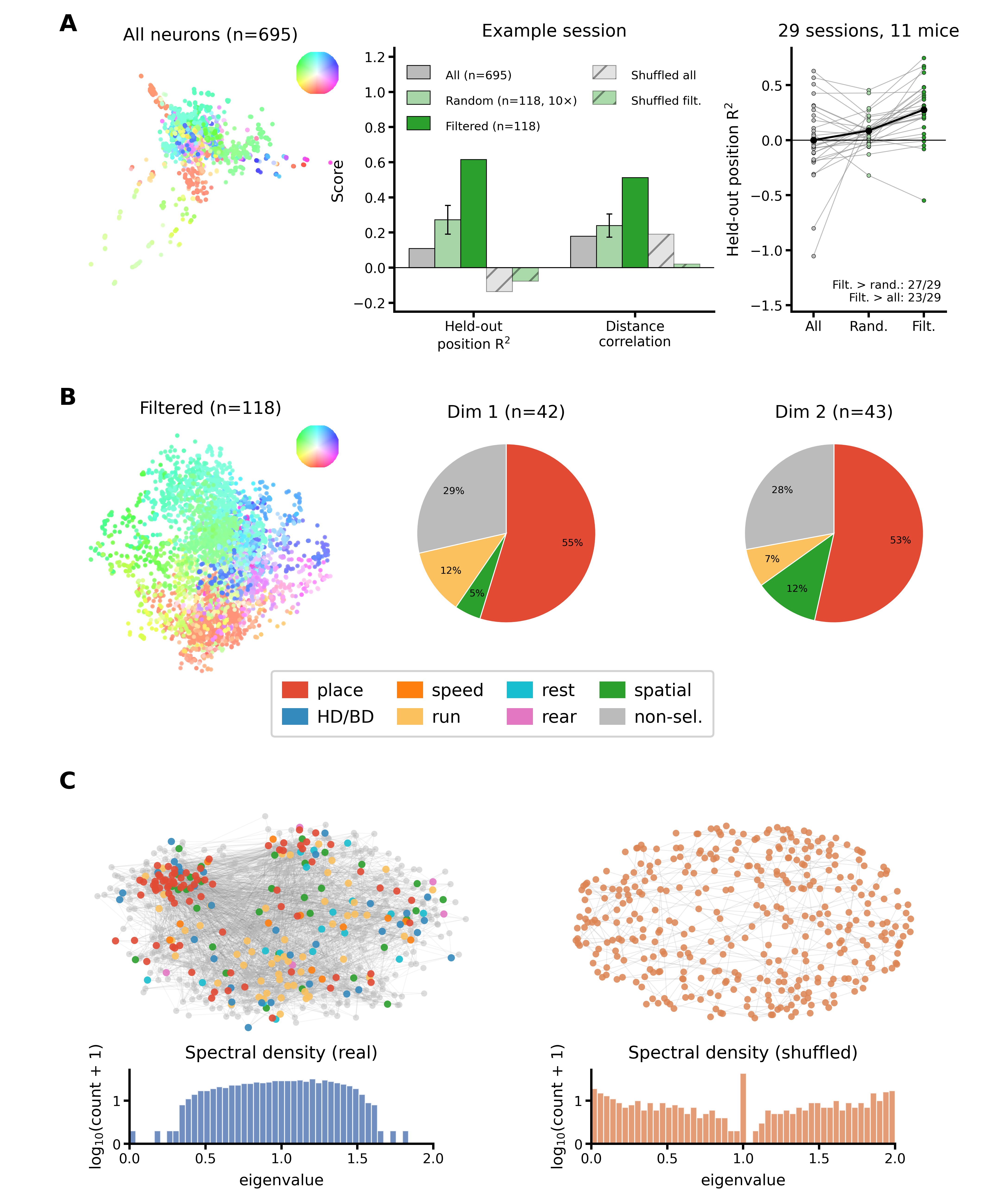}
\caption{\textbf{Scale-bridging integration connects single-neuron
selectivity to population structure and functional networks.}
\textbf{(A)}~Isomap embedding of all 695~neurons,
Procrustes-aligned to physical coordinates and colored by
two-dimensional arena position (left).  Middle: held-out position
$R^2$ and pairwise distance correlation~$\rho$ in this session for
all neurons (gray), 10 random subsets of the filtered-set size (light
green, mean with 95\% CI), spatially filtered
neurons (green, $n = 118$) and circular-shift shuffled controls
(hatched).  Right: held-out position $R^2$ in 29~sessions from
11~mice for all neurons, random subsets (mean per session) and
filtered neurons; gray lines connect values from one session, black
line shows the mean.
\textbf{(B)}~Filtered Isomap embedding ($n = 118$ spatially selective
neurons, left).  Pie charts (right) show the behavioral selectivity
composition of neurons significantly encoding each Isomap dimension, grouped by primary INTENSE
selectivity type.
\textbf{(C)}~Functional network from cell--cell MI significance
testing: real network (left; 12
degree-one nodes omitted from the drawing) with nodes colored by
primary selectivity type (groups as in Methods; ``spatial'' denotes wall or center zone), and shuffled control (right).  Bottom: normalized Laplacian eigenvalue histograms (log$_{10}$ of count + 1); in
the real network two small non-trivial eigenvalues lie below the bulk
(weakly coupled modules), whereas in the shuffled network
the small eigenvalues merge continuously into the bulk, with a sharp
peak at~1 and weight near the bipartite limit of~2, typical of
sparse tree-like graphs.}
\label{fig:scalebridging}
\end{fullwidth}
\end{figure}

\textbf{Selectivity-filtered dimensionality reduction}
(Figure~\ref{fig:scalebridging}A).  The analysis was run on every
session; we first describe the example session.  We embedded the full population
of its 695~neurons and a filtered subset of 118~spatially selective neurons
(identified by INTENSE as carrying significant information about position, wall-zone, or
center-zone occupancy) via Isomap (parameters in Methods) and Procrustes-aligned both embeddings to the animal's
physical coordinates.  Two spatial quality metrics quantify the
effect of filtering (Figure~\ref{fig:scalebridging}A, middle):
position prediction on held-out data (linear regression from
embedding coordinates to position, cross-validated over contiguous
time blocks; Methods) and the correlation between
pairwise distances in the embedding and in the arena.  To separate
the choice of neurons from their number, we compared the filtered
embedding with embeddings of 10 random subsets of 118~neurons.  In
this session, held-out $R^2$ was 0.62 for the filtered embedding and
0.27~$\pm$~0.08 (across the 10 subsets) for random
subsets; distance correlation was $\rho = 0.51$ and
0.24~$\pm$~0.07, respectively.  For the all-neuron embedding,
held-out $R^2$ was at most 0.11 (this value varied between runs)
and distance correlation was 0.18.  In shuffled controls, each neuron's
calcium trace is shifted independently by a circular shift, which
destroys neuron--behavior coupling but keeps single-cell statistics.
They gave negative held-out $R^2$ ($-0.14$ all neurons, $-0.08$
filtered) and $\rho = 0.19$ for all neurons and $0.02$ for the filtered set, so the spatial
structure of the filtered embedding is not explained by single-cell
statistics.

Across all 29~sessions with at least 10~spatially selective neurons
(Methods), the filtered embedding predicted held-out position better than random
subsets of the same size in 27 of 29~sessions and in 10 of 11~mice
($R^2 = 0.28 \pm 0.10$ vs.\ 0.09~$\pm$~0.06; paired Wilcoxon test across sessions,
$p < 0.001$) and better than its shuffled control in 27
of 29~sessions ($-$0.07~$\pm$~0.02)
(Figure~\ref{fig:scalebridging}A, far right).  The two
exceptions were a session with only 11~spatially selective neurons and
one in which the two-dimensional embedding followed time and running
speed rather than position, although place-selective neurons were present.
It also exceeded the
all-neuron embedding ($R^2 = 0.00 \pm 0.14$) in 23 of 29~sessions
($p = 0.001$), but less consistently across animals: in 9 of
11~mice (Wilcoxon test on per-mouse means, $p = 0.02$; sign test,
$p = 0.07$).  Held-out $R^2$ of the filtered embedding increased with the
number of spatially selective neurons (Spearman $\rho = 0.74$ across
29~sessions, $p < 0.001$); it is a linear read-out of an unsupervised
embedding and serves to compare neuron sets rather than to decode
position.  Distance correlation was also higher for filtered
embeddings ($\rho = 0.33 \pm 0.07$) than for random subsets (0.16~$\pm$~0.03; 28 of
29~sessions) and shuffled controls (0.04~$\pm$~0.01; 28 of 29~sessions), and
the all-neuron embedding lay in between (0.20~$\pm$~0.06).

\textbf{Embedding selectivity analysis}
(Figure~\ref{fig:scalebridging}B).  DRIADA can identify which
neurons are significantly informative about each embedding dimension by applying
the same MI-based significance testing used for behavioral features.
Of 695~neurons, 84 significantly encode at least one of the two
dimensions of the filtered embedding ($p < 0.001$; Methods); about 1.4
would be expected by chance at this threshold ($695 \times 2 \times 0.001$).
Grouping these embedding-informative neurons by their primary INTENSE
behavioral selectivity type shows that both selective and nominally non-selective neurons carry information about the embedding: in this session, 24 of the 84 neurons (29\%) were non-selective, compared with 67\% of the population.  Across 29~sessions, 46~$\pm$~5\% of embedding-informative neurons were classified as
non-selective by INTENSE.  This is fewer than the 76~$\pm$~3\% of non-selective
neurons in the same populations (lower in 28 of 29~sessions).  This analysis bridges scales in the
reverse direction, from population structure back to individual
neuron roles, and shows that the coordinates of the spatial population manifold are reflected in the activity of both overtly selective and nominally non-selective neurons.

\textbf{Functional network structure}
(Figure~\ref{fig:scalebridging}C).  We used DRIADA's network module
to construct a functional connectivity graph from pairwise cell--cell
MI significance testing.  The resulting network contains 690~neurons
in the giant connected component with about 3{,}500 significant edges.
In a community-aware spring layout (Louvain communities;
within-community edges weighted 1, between-community edges 0.01), each
node is colored by its primary selectivity type. Place- and run-selective neurons form visually distinguishable
clusters in the layout, indicating that functional connectivity in CA1
reflects shared feature tuning.  Consistent with this, edges
preferentially link neurons of the same primary selectivity type
(attribute assortativity $r = 0.13$, and $r = 0.24$
among selective neurons only; both $p < 0.001$, label
permutation test).  This preference is not explained by selective neurons
having more connections: it persists when labels are permuted only
among neurons of similar degree ($p < 0.001$).  A shuffled-data
control network, constructed from temporally shifted traces, is
substantially sparser (513 edges vs.\ about 3{,}500; its giant
component, shown in the figure, has 361~nodes and 392~edges).  The normalized Laplacian spectral densities
(Figure~\ref{fig:scalebridging}C, bottom) quantify this structural
difference.  In the real network, two small
non-trivial eigenvalues lie below a continuous
bulk.  Such isolated ``soft'' modes indicate a few weakly coupled
modules.  The absence of eigenvalues near~2 indicates the absence of
bipartite substructures.  In the shuffled network, the small
eigenvalues merge continuously into the bulk with no gap, so no
distinct modules can be identified.  Instead, its spectrum shows a
sharp peak at~1, related to topologically equivalent
nodes (neurons with indistinguishable connectivity patterns;
\citealp{Bobyleva2025}) and substantial weight near the bipartite
limit of~2, as expected for a sparse, tree-like graph.

Taken together, these results illustrate DRIADA's central design
principle: the shared data model enables direct transitions between
analysis scales.  Selectivity testing (single neuron) filters and
reshapes dimensionality reduction (population), embedding coordinates
become features for reverse selectivity analysis, and pairwise MI
significance testing yields functional connectivity aligned with
selectivity types (network).

\subsection{Nonlinear manifold structure in a toroidal attractor network}
\label{sec:results-cann}

\begin{figure}
\begin{fullwidth}
\includegraphics[width=0.85\linewidth]{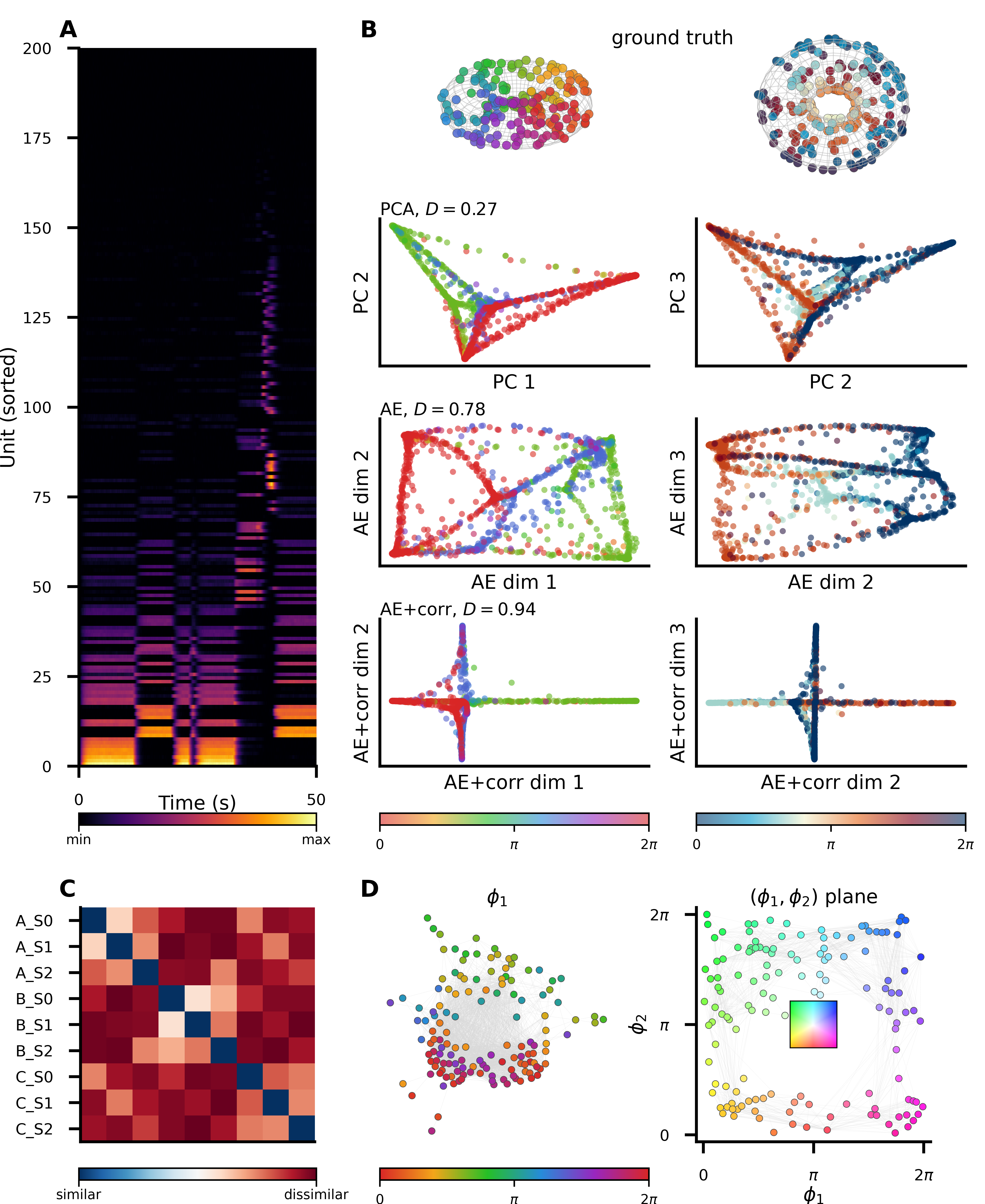}
\caption{\textbf{Multiple DRIADA modules recover the two angular variables of a continuous attractor network on a torus.}
\textbf{(A)}~Activity raster of a 200-unit CANN on $T^2$ driven by
two circular variables $\theta_1$ and $\theta_2$ (first 50~s; units sorted by mean
activity).
\textbf{(B)}~Ground-truth torus (top; neurons at their preferred positions, colored by
$\varphi_1$, left, and $\varphi_2$, right) and two-dimensional projections of the
three-dimensional embeddings: dimensions 1--2 colored by ground-truth $\theta_1$ (left
column) and dimensions 2--3 colored by $\theta_2$ (right column).  PCA
partially entangles the two circular variables; a standard AE
partially disentangles them; an AE with decorrelation loss reaches the highest disentanglement score ($D$ is given above each panel).  The autoencoder embeddings
shown are the best of the 12 initializations.
\textbf{(C)}~Representational dissimilarity matrix for $3 \times 3$
conditions on the torus (A--C: $\theta_1$ position; S0--S2: $\theta_2$ position),
showing block-diagonal structure
consistent with the toroidal geometry.
\textbf{(D)}~Network-of-networks analysis: per-neuron recurrence
graphs combined via Jaccard similarity.  Left: spring layout of the largest
component (rescaled for display, see Methods), colored by $\varphi_1$.  Right: the same
network drawn at the neurons' ground-truth preferred positions in the $(\varphi_1, \varphi_2)$
plane (edges crossing the periodic boundary not drawn).  In both, connected neurons have
nearby preferred positions.}
\label{fig:cann}
\end{fullwidth}
\end{figure}

To test whether DRIADA recovers nonlinear manifold structure,
we simulated a 200-unit continuous attractor neural network (CANN) on a
two-dimensional torus $T^2 = S^1 \times S^1$ (see Methods).  Each neuron
was assigned a preferred position $(\varphi_1, \varphi_2)$ on the torus
with von~Mises recurrent connectivity; two independent circular input
variables $\theta_1(t)$ and $\theta_2(t)$ drove a localized activity
bump across the network
(Figure~\ref{fig:cann}A).  Toroidal CANNs are a standard model of
grid cell population dynamics \citep{Gardner2022} and have been
proposed to underlie three-dimensional head-direction coding in bats,
where azimuth and pitch jointly span a torus
\citep{Finkelstein2015}. We used this system as a controlled
benchmark because the underlying manifold topology is known exactly.

\textbf{Selectivity analysis.}
The INTENSE pipeline identified all 200 neurons as selective
(Holm-corrected $p < 0.001$; see Methods): 172 significantly encoded
$\theta_1$, 150 encoded $\theta_2$, and 122 encoded both.
Per-neuron mutual information values revealed a continuous spectrum
of relative encoding strength: 84 neurons were $\theta_1$-dominant
(MI ratio $> 0.6$), 59 were $\theta_2$-dominant ($< 0.4$), and 57
were balanced, reflecting the
diversity of preferred positions on the torus.

\textbf{Disentangling circular variables via autoencoder embeddings}
(Figure~\ref{fig:cann}B).
The toroidal manifold is difficult for linear
dimensionality reduction: each circular variable requires two linear
coordinates ($\cos\theta$, $\sin\theta$) for faithful representation,
so principal component analysis (PCA) with three components cannot capture both angles
simultaneously \citep{Chung2023}.  We compared three-dimensional
embeddings from PCA, a standard autoencoder (AE), and an autoencoder
with a decorrelation loss (AE+corr) using a disentanglement score $D$ adapted from \citet{Eastwood2018}, which measures whether each latent
dimension encodes a single ground-truth variable ($D = 1$) versus
entangling multiple variables ($D = 0$).

PCA achieved $D = 0.27$, reflecting the mismatch between
linear coordinates and circular topology: each principal component
partially correlates with both angles, as expected from the
cos/sin projections.  A standard autoencoder reached $D = 0.78$
(best of 12 random initializations; mean 0.42~$\pm$~0.12), showing that nonlinear dimensionality reduction can
partially disentangle the two angles.
Adding a decorrelation loss raised the ceiling to $D = 0.94$
(best of 12 initializations; mean 0.38~$\pm$~0.15); both autoencoders were sensitive to initialization.
In the best embedding, the value of each latent dimension followed one of the two circular variables.

\textbf{Representational similarity analysis}
(Figure~\ref{fig:cann}C).
Condition-averaged population vectors were computed for nine
conditions on the torus ($3 \times 3$ grid of $\theta_1$ and
$\theta_2$ positions).  The resulting representational dissimilarity
matrix (RDM; correlation distance) shows block-diagonal structure:
conditions sharing a $\theta_1$ or $\theta_2$ value cluster together, providing a complementary, condition-level view of the same toroidal
geometry recovered by the embedding analysis.

\textbf{Temporal dynamics via recurrence networks}
(Figure~\ref{fig:cann}D).
Per-neuron recurrence graphs were constructed from delay-embedded
activity traces (automatic embedding dimension and delay estimation,
nearest-neighbor thresholding; Methods).  We summarized each neuron by its own
recurrence graph and linked neurons by the similarity between these
graphs, forming a \emph{network of networks}.  Pairwise Jaccard similarity
between recurrence graphs yielded a neuron--neuron functional network,
thresholded to retain the densest connections; its giant connected
component was retained for visualization (Methods).  Spring-layout visualization of this
network orders neurons by their preferred position $\varphi_1$, and when the network is
drawn at the ground-truth preferred positions, its edges connect mainly neurons with
nearby preferred positions: pairwise Jaccard
similarity between recurrence graphs correlated with toroidal
proximity between neurons' preferred positions (Spearman
$\rho = 0.25$ across all 200 neurons; significance assessed by a Mantel permutation test
that respects the non-independence of neuron pairs, $p < 0.001$;
see Methods). The recurrence-derived network
thus reflects the underlying toroidal geometry without access to stimulus
variables.

The four modules thus captured different aspects of the same structure: selectivity testing identifies the continuous gradient of relative encoding strength across the population; the autoencoder assigns the two circular variables to different latent dimensions, which PCA cannot do; RSA shows block-diagonal condition-level geometry; and recurrence networks recover the
spatial organization of the network from temporal dynamics alone,
without access to the stimulus variables.

\section{Discussion}

The four demonstrations show that, with a shared data model, cross-scale analyses can be run without custom pipelines.  On synthetic
data with known ground truth, the end-to-end pipeline recovered
planted selectivity structure, circular manifold geometry, and modular
network organization from the same population
(Figure~\ref{fig:synthetic}).  On hippocampal calcium imaging
recordings from 13~mice, DRIADA detected widespread selectivity across
all 11~behavioral features and characterized the broader selectivity
profile of CA1, including 14\% of selective neurons selective to two or more features
(Figure~\ref{fig:realdata}).  The scale-bridging analysis, repeated
across sessions, showed that in most sessions embeddings of
selectivity-filtered neurons predict held-out position better than
both equally sized random subsets of neurons and the full population,
and that reverse selectivity analysis
identifies the neurons informative about each manifold dimension,
many of which are classified as non-selective by INTENSE
(Figure~\ref{fig:scalebridging}).  On a toroidal continuous
attractor network, four modules (selectivity testing, autoencoder
embedding, representational similarity analysis, and recurrence
networks) gave consistent descriptions of the same two-variable structure
(Figure~\ref{fig:cann}).  DRIADA additionally implements
visibility graphs and ordinal partition networks, addressing  a software
gap identified by \citet{varley2022network}.  We focused the present
demonstrations on recurrence networks because they naturally connect
to the population-level network-of-networks analysis.  Individual
modules have been applied in prior work to hippocampal place-cell
dynamics \citep{Sotskov2022}, population dimensionality estimation
\citep{Pospelov2024}, fMRI resting-state analysis
\citep{Pospelov2021}, structural connectome characterization
\citep{Bobyleva2025}, and functional network entropy
\citep{Pospelov2022}. DRIADA brings these methods into a single integrated framework.

One cross-scale finding is that 46\% of neurons
informative about the spatial population manifold are classified as
non-selective by INTENSE.  This observation reframes ``non-selective'' as a property of the
statistical test and the chosen feature set  rather than the neuron: these cells may carry
sub-threshold spatial tuning too weak to reach significance in
pairwise MI testing, or they may covary with  latent variables not among
the 11~measured features.  \citet{stringer2019high} showed that
population codes in visual cortex are inherently high-dimensional. Our finding is consistent:
neurons that do not individually pass a significance test nonetheless carry information about the spatial manifold.  The scale-bridging
analysis illustrates the scale-dependent artifacts described by
\citet{munn2024scaling}: across sessions, the all-neuron embedding
on average predicted held-out position no better than the mean
position, whereas the selectivity-filtered
embedding did, so a researcher using only population-level
dimensionality reduction could draw substantially different conclusions
about manifold structure.  Our finding extends
\citet{vaccari2022single} in both directions: population analysis benefits from characterizing individual neurons, and characterizing individual neurons alone is incomplete, because nominally non-selective neurons carry information about the spatial manifold.

Most neurons selective to at least one feature were selective to
exactly one feature (Figure~\ref{fig:realdata}C), a predominantly single-feature profile, in contrast to
the strong mixed selectivity reported in prefrontal and association cortex
\citep{rigotti2013importance, fusi2016why}. The rank ordering of
feature prevalences was consistent across all 13~mice despite substantial inter-animal variability in absolute prevalence.
Cross-session analysis of the same neurons recorded on three consecutive days reveals that
selectivity labels are largely session-specific, although retained more often than expected by chance.
This is consistent with representational drift in CA1 \citep{ziv2013longterm, Geva2023HippocampalDrift}.

A design consequence of organizing all graph-based analyses around
a single network representation is that spectral decomposition,
community detection, and entropy measures apply uniformly across
scales.  Functional connectivity graphs derived from cell--cell MI
testing (Figure~\ref{fig:scalebridging}C), proximity graphs produced
by graph-based dimensionality reduction (Laplacian eigenmaps,
diffusion maps, Isomap), and per-neuron recurrence graphs
(Figure~\ref{fig:cann}D) all inherit the same analytical toolkit.
This uniformity opens specific cross-scale comparisons, for example testing whether
communities detected in a functional connectivity network correspond
to clusters in the embedding proximity graph, or whether spectral
properties of recurrence networks predict population-level
organization.

The normalized Laplacian spectra in Figure~\ref{fig:scalebridging}C illustrate this: they distinguish the structured real network from the shuffled one in a single summary, complementing the discrete community assignments from Louvain detection.

Recurrence networks preserve temporal structure that correlation-based
functional connectivity collapses.  From a dynamical systems
perspective \citep{vyas2020computation}, recurrence networks capture
attractor geometry through delay embedding
\citep{donner2010recurrence}, making them sensitive to dynamical
regime, a property not captured by correlation-based edges.
The CANN benchmark (Figure~\ref{fig:cann}D) shows this: the recurrence-derived network reflects the
toroidal organization from temporal dynamics alone, without access to
stimulus variables.

All network analyses in DRIADA are pairwise.
\citet{battiston2021physics} argue that higher-order interactions
involving groups of three or more units are pervasive in neural systems;
the planned integration of partial information decomposition
\citep{williams2010nonnegative} would begin to address this
limitation.

The hippocampal and CANN results both demonstrate that independent
analytical perspectives converge on consistent structure when
operating within a shared representation.  In the hippocampal data,
selectivity testing identifies spatially tuned neurons, population
embedding recovers the arena geometry, and functional network analysis finds connectivity organized by selectivity type, with a modular
spectral structure (Figures~\ref{fig:realdata}--\ref{fig:scalebridging}): complementary analytical
routes reaching the same conclusion.  In the CANN, four modules independently recover the two angular variables that the toroidal
connectivity imposes on the population dynamics, illustrating the prediction of
\citet{dubreuil2022role} that connectivity shapes dynamics. Together, these demonstrations
illustrate the argument of \citet{munn2024scaling} that
multi-scale analysis provides convergent evidence that recovered
structure reflects underlying organization rather than method-dependent artifacts.

\subsection*{Limitations}

The default MI estimator (GCMI) provides a lower bound on mutual
information via Gaussian copula transformation
\citep{ince2017statistical}; for non-monotonic relationships it is therefore conservative rather than
invalid. Because significance is
determined by comparing the observed MI to a shuffled null
distribution, not by its absolute magnitude, a neuron whose MI is
underestimated can still be detected as significant whenever the true
value exceeds the null, albeit with reduced statistical power.
The alternative KSG estimator \citep{kraskov2004estimating} captures
arbitrary dependencies at higher computational cost; a systematic
comparison of estimator choice on detection power is provided in \citet{Pospelov2026intense}.

Although the data model is substrate-agnostic (it has been applied to
artificial recurrent neural networks; \citealp{Kononov2025}), the
selectivity-testing pipeline has been primarily validated on
calcium imaging data. Temporal parameters (circular-shift edge
masks, delay optimization window) are calibrated for the slow
dynamics of calcium indicators; application to electrophysiology or LFP data may require adjustment of these defaults.

The pairwise cell--cell MI computation scales as $O(N^2)$ in neuron
count; for recordings of several thousand neurons, cell--cell significance testing becomes the most expensive step.  The
neuron--feature pipeline (INTENSE) scales linearly in neuron count.

All analyses
operate in batch mode; online or streaming processing is not
currently supported.

The demonstrations presented here show that selectivity type organizes
functional connectivity (Figure~\ref{fig:scalebridging}C), that single-neuron
and population analyses complement each other
(Figure~\ref{fig:scalebridging}A--B), and that temporal dynamics recover the arrangement of neurons by their preferred positions on the torus
(Figure~\ref{fig:cann}D). Each of these cross-scale questions previously required a custom pipeline; with DRIADA's shared data model, each is answered with standard operations.

\clearpage
\section{Methods and Materials}

\begin{fullwidth}
\noindent\textbf{Key resources table}

\smallskip
\noindent{\small
\begin{tabular}{@{}p{3.3cm}p{2.6cm}p{3.6cm}p{2.9cm}p{4.6cm}@{}}
\toprule
Reagent type (species) or resource & Designation & Source or reference & Identifiers & Additional information \\
\midrule
strain, strain background (\textit{Mus musculus}, males and females) & C57BL/6J & Laboratory Animal Breeding Center ``Pushchino'' & & 14 mice used, 13 analysed (5 females, 8 males); see Animals \\
strain, strain background (adeno-associated virus) & AAV/DJ-CAG-GCaMP6s & Federal Medical-Biological Agency of Russia (FMBA) & & 300~nL into dorsal CA1; see Surgeries and viral delivery \\
other & Miniscope V4.4 & Open Ephys & RRID:SCR\_021480 & Head-mounted one-photon miniscope; see Behavioral procedure \\
software, algorithm & DRIADA 1.2.2 & This paper; \citet{driada2026software} & & \url{https://github.com/iabs-neuro/driada/releases/tag/v1.2.2}; see Implementation and availability \\
software, algorithm & BEARMIND & Other & & Calcium imaging pipeline built on CaImAn, \url{https://github.com/iabs-neuro/bearmind}; see Processing of calcium imaging data \\
software, algorithm & SPHYNX & \citet{Plusnin1} & & \url{https://github.com/iabs-neuro/sphynx}; see Behavioral analysis \\
software, algorithm & CaImAn & \citet{giovannucci2019caiman} & RRID:SCR\_021152 & \\
software, algorithm & NoRMCorre & \citet{Pnevmatikakis2017} & & Motion correction \\
software, algorithm & CNMF-E & \citet{Zhou2018} & & Source extraction \\
software, algorithm & DeepLabCut & \citet{Mathis} & RRID:SCR\_021391 & Pose estimation \\
software, algorithm & Bonsai & Bonsai Foundation & RRID:SCR\_017218 & Synchronization of behavioral and miniscope streams \\
\bottomrule
\end{tabular}}
\end{fullwidth}

\subsection{Unified data representation}

All modules operate on a shared hierarchy of data objects that
eliminates format conversion between analysis stages.  The top-level
container, \texttt{Experiment}, stores a calcium-activity matrix of
shape $(N_\text{cells}, N_\text{frames})$, an optional spike matrix of
equal shape, a dictionary of time-aligned behavioral variables
(\texttt{dynamic\_features}), time-invariant metadata
(\texttt{static\_features} such as frame rate and indicator kinetics),
and per-neuron objects (\texttt{Neuron}).  Each \texttt{Neuron} wraps
its calcium and spike traces as \texttt{TimeSeries} objects; the full
population is jointly accessible as a \texttt{MultiTimeSeries}, which
inherits dimensionality-reduction capabilities from its parent class
\texttt{MVData}.

\texttt{TimeSeries} is the fundamental univariate container.  On
construction, it runs automatic type detection
(\texttt{analyze\_time\_series\_type}) that classifies the signal as
continuous (linear or circular) or discrete (binary, categorical,
count, or timeline) and records the result as a \texttt{TimeSeriesType}
object carrying subtype, confidence, and (for circular
variables) period.  The detection uses statistical heuristics
(uniqueness ratio, integer checking, autocorrelation peak analysis) and
can be overridden by the user via an explicit \texttt{ts\_type}
parameter.  From this classification, \texttt{TimeSeries} lazily
precomputes the representations needed by downstream estimators:
copula-normalized values (\texttt{copula\_normal\_data}) for GCMI
\citep{ince2017statistical}, integer-coded values (\texttt{int\_data})
for discrete entropy, scaled values (\texttt{scdata}) for correlation,
and a KD-tree (\texttt{get\_kdtree}) for KSG
\citep{kraskov2004estimating}.  Shannon entropy values are cached per
downsampling factor to avoid recomputation across modules.

\texttt{MultiTimeSeries} extends this pattern to multiple aligned
series.  It validates that all component \texttt{TimeSeries} share the
same type and length, combines their shuffle masks via logical AND,
and exposes joint entropy and copula-normalized matrices.  Because
\texttt{MultiTimeSeries} inherits from \texttt{MVData}, any neural
population matrix can be reduced to a low-dimensional embedding via
\texttt{get\_embedding(method, dim)} without leaving the data model.
Graph-based methods (Laplacian eigenmaps, diffusion maps, UMAP,
Isomap) additionally return a \texttt{ProximityGraph} that inherits
the full network-analysis toolkit (see Network analysis below).

When an \texttt{Experiment} is constructed, behavioral variables are
automatically wrapped as typed \texttt{TimeSeries} with appropriate
shuffle masks that exclude short circular shifts, preventing edge
artifacts during circular-shift permutation testing: shifts within
$2$~s of zero are excluded for behavioral features and shifts within
five calcium decay constants (10~s by default) for calcium traces; the
two masks are combined for each pair.  For circular features such as head direction, the framework
automatically constructs a two-dimensional $(\cos,\sin)$
representation, enabling MI estimation that respects the circular
topology.  Multivariate features (e.g., two-dimensional position) are
stored as \texttt{MultiTimeSeries}, and the same MI core handles
them via multivariate estimators.  This automatic variable handling
means that a single call to the selectivity-testing pipeline
processes all features regardless of type, with no manual configuration.

\subsection{Information-theoretic selectivity testing}

DRIADA integrates the INTENSE algorithm for detecting statistically
significant neuron--feature associations; its algorithmic details,
including the zero-inflated gamma null model and the fast Fourier
transform (FFT) acceleration, are described in
\citet{Pospelov2026intense}.  In this work,
\texttt{compute\_cell\_feat\_significance()} tested every
neuron--feature pair in two stages: a screen with 100 circular time
shifts, followed, for pairs that passed it, by a validation stage with
$\min(10{,}000,\, N_{\mathrm{ds}})$ shifts, where $N_{\mathrm{ds}}$ is
the number of samples after downsampling, and a $p$-value from a
zero-inflated gamma fit to the null MI values.  Pairs were called
significant after Holm--Bonferroni correction at $\alpha = 0.01$ for the
synthetic data (Figure~\ref{fig:synthetic}) and at a per-pair threshold
of $p < 0.001$ without family-wise correction for the hippocampal
dataset (Figures~\ref{fig:realdata} and \ref{fig:scalebridging}; threshold
selection is discussed in \citet{Pospelov2026intense}); the settings for
Figure~\ref{fig:cann} are given with the CANN simulation below.

Circular time shifts, unlike naive shuffles, preserve the temporal
autocorrelation of both calcium and behavioral signals, and MI at all
shifts is computed at once with the FFT.  For 500 neurons and 20 binary
features in a 1-h recording at 20~Hz, the two-stage test takes about one
minute on a laptop CPU (Intel Core Ultra 9 288V, single thread).

Two mutual-information estimators are available.  The default, GCMI
\citep{ince2017statistical}, transforms signals to Gaussian copula
space and computes MI via a closed-form expression that integrates
directly into the FFT pipeline; it captures monotonic nonlinearities
and is fast enough for large datasets.  The KSG estimator
\citep{kraskov2004estimating} uses $k$-nearest-neighbor statistics and
captures arbitrary nonlinear dependencies, including bell-shaped and
ROI-based tuning, at higher computational cost.  GCMI is used by
default for all variable types (only its variant depends on the
variable type); KSG must be requested explicitly.

Delay optimization searches within a window of $\pm 60$ frames in
five-frame steps ($\pm 2$~s in the hippocampal recordings) to account for GCaMP indicator kinetics and prospective or
retrospective encoding.  An optional disentanglement stage uses
conditional MI and interaction information to separate genuine mixed
selectivity from spurious associations caused by covarying behavioral
variables; in this paper it was used only for the counts of
Figure~\ref{fig:realdata}C.

Selectivity in the hippocampal recordings was computed with type-based
feature representations, the default in DRIADA~1.2.1: speed entered the
MI estimator as $[x, (x-c)^2]$ with $c$ its median; head and body
direction as $[\cos\theta, \sin\theta,$ $\cos 2\theta, \sin 2\theta]$;
position as $[x, y,$ $(x-c_x)^2, (y-c_y)^2,$ $(x-c_x)(y-c_y)]$; discrete
states unchanged.  Pairs in which the neuron was less active when a
binary state was present (or, for speed, above its median) than
otherwise were not counted as selective (2{,}410 pairs).

\begin{sloppypar}
The same circular-shift MI framework extends to two additional
testing modes.  \texttt{compute\_cell\_cell\_significance()} tests all
neuron--neuron pairs rather than neuron--feature pairs, producing a
binary significance matrix that serves as the adjacency matrix for
functional connectivity networks (Figures~\ref{fig:synthetic}D--E,
\ref{fig:scalebridging}C).  \texttt{compute\_embedding\_selectivity()}
treats embedding coordinates (e.g., Isomap or UMAP dimensions) as
continuous features and applies the same two-stage permutation test to
identify which neurons significantly encode each embedding dimension
(Figure~\ref{fig:scalebridging}B).  Both functions inherit the
circular-shift permutation scheme, FFT acceleration, and optional
multiple-comparison correction from the core pipeline; neuron--neuron
testing uses no delay optimization (zero lag in both stages), and its
validation stage used 1{,}000 shifts.
\end{sloppypar}

Shuffled-data controls were built by circularly shifting each neuron's
calcium trace by an independent random offset (Figure~\ref{fig:realdata}A:
between 1{,}000 and $N - 1{,}000$ frames; Figure~\ref{fig:scalebridging}A:
within the central 80\% of the downsampled recording;
Figure~\ref{fig:scalebridging}C: no shorter than three decay time
constants), which preserves each neuron's temporal autocorrelation and
amplitude distribution while destroying inter-neuron and
neuron--behavior relationships.

\subsection{Dimensionality reduction}

The dimensionality-reduction module provides 15 methods accessible
through a unified interface
(\texttt{MVData.get\_embedding(method, dim)}).  Methods span five
families: linear (PCA), spectral graph-based (Laplacian eigenmaps,
diffusion maps \citep{coifman2006diffusion}, Isomap, locally linear
embedding, Hessian LLE, maximum variance unfolding), distance-based
(MDS), neighborhood-based (t-SNE, UMAP \citep{mcinnes2018umap}), and
neural network-based (autoencoder, variational autoencoder, and a
user-configurable flexible autoencoder).  Library variants are
available for Laplacian eigenmaps (\texttt{auto\_le}, scikit-learn
spectral embedding) and diffusion maps (\texttt{auto\_dmaps}, pydiffmap
with automatic bandwidth selection).  All methods return an \texttt{Embedding} object that
stores coordinates, the fitted model, and (for graph-based
methods) a \texttt{ProximityGraph} inheriting the full
network-analysis toolkit (see Network analysis below).

Because neural activity matrices are stored as
\texttt{MultiTimeSeries}, which inherits from \texttt{MVData},
population data can be embedded directly.  Graph-based methods first
build a $k$-nearest-neighbor graph (default $k = 15$; $k = 10$ for LLE
and Hessian LLE; Euclidean metric; approximate search), symmetrized by
union (two points are joined if either
is among the $k$ nearest neighbors of the other) and, except for UMAP,
restricted to its giant connected component.  Graphs are unweighted
except for diffusion maps, which use heat-kernel affinities.  Laplacian
eigenmaps take the leading non-trivial eigenvectors of the graph's
random-walk transition matrix, and diffusion maps those of the
$\alpha$-normalized diffusion operator ($\alpha = 0.5$); Isomap applies
kernel PCA to shortest-path (hop-count) distances on the graph; UMAP and
LLE delegate the embedding to umap-learn and scikit-learn with the same
$k$, each building its own neighbor graph.  The DRIADA graph is retained
so that subsequent spectral
and community analyses can be performed without graph reconstruction.
Sequential reduction pipelines (e.g., PCA to 50 dimensions followed by
UMAP to 2 dimensions) are supported through
\texttt{dr\_sequence()}.

\begin{sloppypar}
Embedding quality is assessed by several complementary metrics:
\texttt{trustworthiness} and \texttt{continuity}
\citep{venna2006local} measure how well the embedding preserves
local neighborhoods; $k$-nearest-neighbor preservation rate
(\texttt{knn\_preservation\_rate}) quantifies the fraction of
neighbors retained; and geodesic distance correlation evaluates global
structure preservation.  For neural network methods, reconstruction
error and latent-space metrics are reported automatically.
Procrustes analysis (\texttt{procrustes\_analysis}) aligns a
target embedding to a reference coordinate system via rotation and,
optionally, scaling and reflection, enabling direct visual comparison
between neural manifolds and behavioral variables such as spatial
position.  For the analyses in this paper, two additional metrics
were computed in the analysis scripts rather than by the package, using
the animal's physical position as reference coordinates: the
\emph{position $R^2$} (multivariate linear regression from embedding
coordinates to reference position), and the \emph{pairwise distance
correlation} (Spearman $\rho$ between pairwise distances in embedding
space and reference space, subsampled to 2{,}000 points for
tractability).
\end{sloppypar}

The dimensionality-estimation module complements reduction with tools
for determining intrinsic dimensionality: PCA-based thresholding
(fraction of variance explained), effective rank (spectral entropy of
eigenvalues), $k$-nearest-neighbor dimension, correlation dimension
(Grassberger--Procaccia), and geodesic dimension
\citep{granata2016accurate}.  These estimators inform the choice
of target dimensionality for reduction.

\subsection{Network analysis}
\label{sec:methods-network}

The \texttt{Network} class provides a general-purpose graph-analysis
framework that underlies both functional connectivity analysis and
the proximity graphs produced by graph-based dimensionality reduction.
A \texttt{Network} can be constructed from a sparse adjacency matrix
or a NetworkX graph.  Directed, undirected, weighted, and unweighted
graphs are supported; directedness and weights are auto-detected from
the adjacency structure when not specified explicitly.

Spectral decomposition is central to the analysis.  The class computes
eigenvalues and eigenvectors of the adjacency matrix, graph Laplacian,
normalized Laplacian, random-walk Laplacian, or transition matrix on
demand.  From the spectrum, several summary statistics are derived:
the inverse participation ratio (IPR) of each eigenvector quantifies
its localization; free entropy, R\'{e}nyi $q$-entropy, and spectral
entropy characterize the graph's thermodynamic properties at
user-specified diffusion times; and the Estrada communicability index
measures global connectivity.  Gromov hyperbolicity estimates
characterize the global geometry of the graph.  Degree-preserving
randomization generates null models for statistical comparison.
Each \texttt{Network} exposes its NetworkX graph, so community
detection (e.g., Louvain) can be applied directly.
Graph preprocessing options include extraction of the giant connected
or strongly connected component and removal of self-loops and
isolated nodes.

Because \texttt{ProximityGraph} (used by graph-based DR methods)
inherits from \texttt{Network}, any embedding that produces a
proximity graph (Laplacian eigenmaps, diffusion maps, Isomap,
UMAP) automatically exposes the full spectral and community analysis
toolkit. Network analysis of an embedding therefore needs no additional code.

Representational similarity analysis (RSA)
\citep{kriegeskorte2008representational} is provided by a dedicated
module.  \texttt{compute\_rdm()} constructs representational
dissimilarity matrices from neural activity patterns using
correlation, Euclidean, cosine, or Manhattan distance.  Multiple
entry points support different experimental designs:
condition-averaged patterns, trial-level patterns, or labeled time
series.  RDMs are compared via Spearman or Pearson correlation, with
significance and confidence intervals obtained by within-condition bootstrap resampling
(\texttt{bootstrap\_rdm\_comparison}).

\subsection{Network analysis of time series}

Three complementary methods for constructing networks directly from
time series \citep{varley2022network} are implemented as subclasses of
\texttt{Network}, inheriting the full spectral and community analysis
toolkit.

\textbf{Recurrence networks} \citep{donner2010recurrence,
marwan2007recurrence}.  A univariate time series is first embedded in
a reconstructed phase space via Takens time-delay embedding
(\texttt{takens\_embedding}), with embedding dimension $m$ and delay
$\tau$ estimated automatically via false nearest neighbors
\citep{kennel1992determining} and the first minimum of time-delayed
mutual information, respectively.  Pairs of embedded state vectors
closer than a threshold (either $\varepsilon$-ball or $k$-nearest
neighbors) are connected by an edge; an optional Theiler window can
remove trivially correlated temporal neighbors.  Recurrence quantification
analysis (\texttt{compute\_rqa}) extracts determinism, laminarity,
trapping time, and diagonal-line entropy from the recurrence matrix.
Community detection on the recurrence network identifies dynamical
macro-states.  At the population level, per-neuron recurrence matrices
are combined into a joint recurrence plot (JRP) via thresholded
element-wise intersection, and pairwise Jaccard similarity between
individual matrices yields a neuron--neuron network amenable to
further community analysis.

\textbf{Visibility graphs} \citep{lacasa2008time,
luque2009horizontal}.  Two time points are connected if no
intermediate value obstructs the ``line of sight'' between them.  The
horizontal visibility graph (HVG) uses an $O(N)$ monotone-stack
algorithm and requires no free parameters; the natural visibility
graph (NVG) uses an $O(N^2)$ pairwise check and captures additional
geometric structure.  The degree distribution of the resulting graph
reflects the dynamical regime: fractal processes produce scale-free
distributions, while uncorrelated random processes yield exponential
distributions.  Hub nodes correspond to extreme events in the
original signal.

\textbf{Ordinal partition networks} \citep{bandt2002permutation,
mccullough2015time}.  The time series is embedded via Takens
delay embedding and each state vector is reduced to its rank-order
pattern (ordinal pattern), encoded as a Lehmer code (an integer in
$[0,\, d!)$, where $d$ is the embedding dimension).
Nodes of the network represent unique ordinal patterns; directed
weighted edges encode transition probabilities between consecutive
patterns (row-normalized, self-loops removed).  Permutation entropy
(the normalized Shannon entropy of pattern visit frequencies)
summarizes the complexity of the dynamics on a $[0,1]$ scale.
Missing patterns indicate forbidden ordinal transitions.

\begin{sloppypar}
All three graph types are accessible through the \texttt{TimeSeries}
API (\texttt{ts.recurrence\_graph()},
\texttt{ts.visibility\_graph()},
\texttt{ts.ordinal\_partition\_network()}) with automatic parameter
estimation and result caching.
\end{sloppypar}

\subsection{Synthetic data generation}

DRIADA includes synthetic data generators that produce populations of
neurons with known selectivity profiles, enabling quantitative
validation of the full analysis pipeline.  The canonical generator
(\texttt{generate\_tuned\_selectivity\_exp}) takes a declarative
population specification: each neuron group is defined by its count,
the features it responds to, and the tuning curve type: von Mises for
head-direction selectivity, two-dimensional Gaussian for place fields,
sigmoid for speed tuning, and threshold activation for discrete
events.  Multiple selectivity can be specified via OR (responsive to
any feature) or AND (responsive only when all features are active)
combination modes.

Behavioral variables are generated as correlated random processes:
continuous features use fractional Brownian motion (fBM) with a
configurable Hurst exponent $H$ (default $H = 0.3$, producing
anti-persistent trajectories with naturalistic episode structure);
two-dimensional position follows a momentum-driven random walk with
boundary reflection; circular variables (head direction) follow a
circular random walk; and discrete events are generated as binary
time series with configurable active fraction and average episode
duration.

Neural responses are converted to realistic calcium fluorescence
traces by convolving spike trains (generated via an inhomogeneous
Poisson process with tuning-modulated rates) with a
single-exponential kernel (instantaneous rise, decay time constant
2.0~s).  A double-exponential kernel with a separate
rise phase is available as an option but was not used here.
Additive Gaussian noise, random event amplitudes, and a configurable
skip probability (probability of failing to respond during an active
period) allow systematic control of the signal-to-noise ratio.

The generator returns a standard \texttt{Experiment} object with
embedded ground truth (expected neuron--feature pairs, neuron type
labels, and tuning parameters), so that detection rates (TPR, FPR,
F1) can be computed directly against the known selectivity structure.

\begin{sloppypar}
For Figure~\ref{fig:synthetic}, the event-modulated and head-direction
populations were simulated as two independent 600-s recordings at
20~Hz and analyzed separately.  Three binary event features were
generated with a target active fraction of 8\% and mean episode
duration of 0.8~s (realized 7--8\% and 0.73--0.85~s); event neurons
fired at 2~Hz during their events and 0.05~Hz otherwise, with no skipped responses.
Head-direction neurons were generated with a dedicated generator
(\texttt{generate\_circular\_manifold\_exp}): head direction follows a
circular random walk, preferred directions are evenly spaced with small
jitter, and rates range from 0.1~Hz baseline to 2~Hz peak (von~Mises
tuning, $\kappa = 4$).  All analyses used every fifth frame, and Isomap
used $k = 50$ neighbors.  After aligning the embedding angle to head
direction by the rotation and reflection that minimize the mean absolute
angular error, agreement was quantified as $\rho_{\mathrm{circ}}$, the
modulus of the complex correlation between the mean-centered unit
phasors of the two angle series.  Its significance was assessed by
circularly shifting the head-direction trace by a uniformly random
offset of 1 to $n - 1$ samples (1{,}000 shifts), refitting the alignment
each time; $p = (k + 1)/1{,}001$, where $k$ is the number of shifts with
$\rho_{\mathrm{circ}}$ at least as large as the observed one.  An edge
of the network in Figure~\ref{fig:synthetic}D--E is a neuron pair
significant in the two-stage test (100 and 1{,}000 shifts, Holm
correction at 0.01).  ARI and NMI were computed over all 300 neurons
between the seven planted groups and the detected partition (three
communities plus one class for neurons outside the giant component), so
the maximum attainable agreement is below 1.
\end{sloppypar}

\subsection{Continuous attractor network simulation}
\label{sec:methods-cann}

The toroidal CANN benchmark (Figure~\ref{fig:cann}) simulates a
rate-model network of $N = 200$ units on a two-dimensional torus
$T^2 = S^1 \times S^1$.  Each neuron~$i$ is assigned a uniformly
random preferred position $(\varphi_1^{(i)}, \varphi_2^{(i)}) \in
[0, 2\pi)^2$.  The membrane potential $x_i$ evolves as
\begin{equation}
\tau \frac{dx_i}{dt} = -x_i + \sum_j W_{ij}\, r_j
  + I_{\mathrm{ext},i}(t) + \eta_i(t)
\end{equation}
with firing rate $r_i = [x_i]_+$, ReLU activation
$[\cdot]_+ = \max(\cdot, 0)$, time constant $\tau = 0.2$~s, and Gaussian
noise $\eta_i$ drawn independently from $\mathcal{N}(0, 0.05^2)$ at every
step; the equation was integrated by the forward Euler method with step
$\Delta t = 0.05$~s from $x_i(0) = 0$.  Recurrent weights follow a
translation-invariant von~Mises profile with global inhibition:
\begin{equation}
W_{ij} = \frac{J_\mathrm{exc}}{N}\,
\exp\!\Big(\kappa_c\big(\cos(\varphi_1^{(i)} - \varphi_1^{(j)})
+ \cos(\varphi_2^{(i)} - \varphi_2^{(j)}) - 2\big)\Big)
- \frac{J_\mathrm{inh}}{N}
\end{equation}
with $J_\mathrm{exc} = 8$, $J_\mathrm{inh} = 0.8$, and connectivity
concentration $\kappa_c = 2.0$.  External input positions a bump at
$(\theta_1(t), \theta_2(t))$ via
$I_{\mathrm{ext},i} = \exp(\kappa_\mathrm{in}(\cos(\theta_1 - \varphi_1^{(i)})
+ \cos(\theta_2 - \varphi_2^{(i)}) - 2))$
with input sharpness $\kappa_\mathrm{in} = 4.0$.

Two independent circular latent variables $\theta_1(t)$ and
$\theta_2(t)$ are generated as switching processes: each jumps
between three equally-spaced positions on $[0, 2\pi)$ with
exponentially distributed dwell times (mean 10~s and 6~s
respectively) and smooth transitions ($\tau_\mathrm{tr} = 0.3$~s).
The simulation runs for 600~s at 20~Hz (12{,}000 frames).  Raw
firing rates are exponentially smoothed ($\tau_\mathrm{smooth} = 0.5$~s)
and corrupted with additive observation noise
($\sigma_\mathrm{obs} = 0.01$) to mimic calcium indicator dynamics.

Selectivity in the CANN was tested with the two-stage procedure (GCMI;
100 and then 2{,}400 circular shifts, that is $\min(10{,}000,\, N_{\mathrm{ds}})$;
Holm correction at $\alpha = 0.001$)
on activity downsampled five-fold, with each circular variable
represented as $[\cos\theta, \sin\theta, \cos 2\theta, \sin 2\theta]$
(the type-based representation, as for head direction in the
hippocampal data).  For each neuron the MI ratio
was defined as $\mathrm{MI}(\theta_1)/(\mathrm{MI}(\theta_1) +
\mathrm{MI}(\theta_2))$, computed for all 200 neurons irrespective of
significance.

For disentanglement analysis, three-dimensional embeddings were
computed via PCA and two autoencoder variants.  The encoder and
the decoder each have one hidden layer of 128 units with leaky-ReLU
activation; the 3-dimensional bottleneck and the output layer are
linear.  The standard AE minimizes reconstruction
error only (MSE loss); the decorrelating variant adds a pairwise
correlation penalty on the latent dimensions (weight 1.0).  Both are
trained for 200 epochs (Adam, $\mathrm{lr} = 10^{-3}$, batch size
512) on the first 80\% of frames, with input dropout ($p = 0.2$; the
reconstruction target is the masked input); the network from the epoch
with the lowest loss on the remaining 20\% of frames (validation set)
was used, and embeddings and $D$ were computed on all
12{,}000 frames.  To account for sensitivity to initialization, each autoencoder
is trained with 12 random seeds; the best $D$ is reported
together with the mean $\pm$ half-width of the 95\% confidence interval across seeds.
Disentanglement is quantified by a score adapted from the
disentanglement score of \citet{Eastwood2018}: for each latent
dimension $z_d$ and each angle $\theta_k$ we compute the $R^2$ of the
least-squares fit $z_d = a\cos\theta_k + b\sin\theta_k + c$; these
importance weights are normalized to a probability distribution over
the ground-truth variables, and $D = 1 - \bar{H} /
\log K$, where $\bar{H}$ is the unweighted mean entropy over the three
dimensions (unlike the importance-weighted mean of the original; a
dimension with summed $R^2 < 0.01$ is assigned the maximal entropy) and
$K = 2$ is the number of ground-truth variables.  $D = 1$ indicates
each dimension encodes exactly one variable; $D = 0$ indicates
maximal entanglement.

For RSA, nine conditions were defined by crossing the three
$\theta_1$ positions with the three $\theta_2$ positions; each frame
was assigned to one of the nine conditions by the pair of current target
positions of $\theta_1$ and $\theta_2$ (all frames, including
transitions), and the
representational dissimilarity matrix was computed from
condition-averaged population vectors using correlation distance.

For the recurrence analysis of Figure~\ref{fig:cann}D, the activity
trace of each of the 200 units was downsampled five-fold (2{,}400
samples) and delay-embedded with a unit-specific delay (first minimum of
the time-delayed mutual information, up to 50 samples) and dimension
(false nearest neighbors, $2 \le m \le 8$).  Each embedded point was
linked to its 20 nearest neighbors (Euclidean distance, union
symmetrization); no Theiler window was applied.  Recurrence matrices
were truncated to the shortest embedded length, and similarity between
two units was the Jaccard index of their edge sets.  For the network
of Figure~\ref{fig:cann}D the Jaccard matrix was thresholded at its
85th percentile (2{,}985 of 19{,}900 pairs), and the largest connected
component (136 neurons, 2{,}969 edges) was drawn with the
Fruchterman--Reingold spring layout; for display, node positions were
radially rescaled ($r \to r^{0.35}$) and rotated by 45 degrees.

The relation between recurrence-graph similarity and tuning proximity
(Figure~\ref{fig:cann}D) was tested on all 200~neurons, that is on
19{,}900 neuron pairs, using the full matrix of pairwise Jaccard
similarities before thresholding.  Toroidal proximity of two neurons
was defined as the negative of the distance
$\sqrt{\Delta_1^2 + \Delta_2^2}$ between their preferred positions,
where $\Delta_1$ and $\Delta_2$ are the circular differences in
$\varphi_1$ and $\varphi_2$, and the test statistic was the Spearman
correlation between Jaccard similarity and proximity over all pairs.
Because each neuron enters 199~pairs, the pairs are not independent
and the usual $p$-value of a correlation does not apply; we therefore
used a Mantel permutation test \citep{mantel1967detection}, in which the neuron labels of the
proximity matrix are permuted (the same random permutation applied to
its rows and columns) and the correlation is recomputed, so that the
dependence among pairs sharing a neuron is preserved under the null
hypothesis.  With $10{,}000$ permutations, the one-sided $p$-value was
$(b + 1)/(10{,}000 + 1)$, where $b$ is the number of permutations
giving a correlation at least as large as the observed one; no
permutation did ($b = 0$), so $p \approx 10^{-4}$, the smallest value
this number of permutations can return.

\subsection{Hippocampal calcium imaging dataset}
\label{sec:methods-fof}

\subsubsection{Animals}
\label{sec:fof-animals}

We used C57BL/6J mice (both sexes, $n = 14$), aged 10--11 months.
Before surgery, animals were housed in standard laboratory cages
(2--7 per cage) under a 12-hour light/dark cycle, with free access
to food and water.  After surgery mice were housed individually;
other housing conditions remained unchanged.  All experiments were
performed during the light phase of the day (10:00 a.m.\ to
6:00 p.m.).  One of the 14 implanted mice was excluded because of
aberrant calcium-wave activity in the imaging field, leaving 13 mice
(5 females and 8 males) in the final analysis; this exclusion criterion
was set before the analysis.
Data from females and males were pooled.

\subsubsection{Surgeries and viral delivery}
\label{sec:fof-surgeries}

All procedures were approved by the Commission of Bioethics of
Lomonosov Moscow State University (Application \textnumero{} 159-g,
approved during Bioethics Commission meeting \textnumero{} 154-d-r
held on 17.08.2023), and followed the Russian Federation Order
N~267~MZ and the NIH Guide for the Care and Use of Laboratory
Animals.

Surgical procedures were carried out in three stages: viral
injection, GRIN lens implantation, and baseplate installation.
Anesthesia was induced by intraperitoneal injection of zoletil
(40~mg/kg) and xylazine (5~mg/kg).  Dexamethasone (4~mg/kg) was
administered subcutaneously 5 minutes before surgery to reduce
inflammation and prevent cerebral edema. Moisturizing gel was applied to prevent eye drying.  Mice were positioned in a stereotaxic
frame (Kopf, USA), and body temperature was maintained with a heated
pad (Physitemp, USA).  For virus injection, 300~nL of
AAV/DJ-CAG-GCaMP6s was delivered unilaterally into the dorsal CA1
region (coordinates: AP $-1.94$~mm, ML $+1.46$~mm, DV $-1.2$~mm;
\citealp{Paxinos2019}).  One week later, a 1-mm-diameter GRIN lens
(GrinTech, Germany) was implanted 200~$\mu$m above CA1.  The lens
was attached to the skull with cyanoacrylate glue (Henkel, Germany)
and dental cement (Stoelting, USA), and covered with Kwik-Sil (World
Precision Instruments, USA).  One
week after lens implantation, an aluminum baseplate was placed over the implant site for miniscope mounting and secured to the skull with dental cement.  A removable
plastic cap was placed over the implant  to protect the lens. Following the final surgery, mice were returned to their home cages to recover.

\subsubsection{Behavioral procedure}
\label{sec:fof-behavior}

Two weeks after the final surgery, mice were gradually habituated to
handling and wearing a miniscope.  For three consecutive days, each
mouse was placed in its home cage with a connected but non-recording
miniscope for 5 minutes per session. On the following three days, animals
were tested in a circular open field arena (diameter 96~cm) with
visual landmarks placed on the surrounding walls. Each session lasted approximately 12 minutes. Throughout each trial, animal behavior was
recorded using a top-mounted video camera (Flir Chameleon3, KVR,
UK), while CA1 calcium activity was simultaneously acquired via the
miniscope (Miniscope V4.4).  Behavioral video and miniscope streams
were synchronized in Bonsai (Bonsai Foundation, UK) using shared
timestamps, yielding aligned time bases for delay-dependent analyses.

\subsubsection{Behavioral analysis}
\label{sec:fof-behavioral-analysis}

Behavioral tracking and annotation were performed using the SPHYNX
package for exploratory behavior analysis \citep{Plusnin1}.  For each
video, the positions of key body parts were estimated frame-by-frame
using a DeepLabCut-trained neural network \citep{Mathis}.  Coordinates
with confidence scores below 0.95 were excluded and reconstructed by
piecewise cubic interpolation using neighboring high-confidence
frames.  The resulting trajectories were smoothed with a third-degree
Savitzky--Golay filter, applied with a 0.25-second window for the
body center and tail base, and 0.1 seconds for all other body parts.

The arena was segmented into functionally distinct spatial zones: a
10-cm-wide wall zone along the perimeter, a 20-cm-wide intermediate
ring, and a central zone (circle of diameter 40~cm).  Both continuous
and discrete behavioral variables were calculated based on the
animal's movement and position relative to these areas.  Continuous
variables included: Cartesian coordinates of the body center ($x$,
$y$); absolute velocity, calculated from body center displacement and
smoothed with a 0.25-second window; body direction~(BD), defined as
the angle from body center to head center; and head direction~(HD),
defined as the angle from head center to nose tip.

Discrete behavioral acts were classified based on movement dynamics.
Locomotor states were defined by current speed: fast locomotion
(run, $> 5$~cm/s), slow locomotion (walk, 1--5~cm/s), and rest ($< 1$~cm/s).
Freezing was defined as simultaneous decrease of both body center
velocity ($< 1$~cm/s) and nose tip velocity ($< 2$~cm/s).  Rearing
was detected by a transient reduction in the distance between the
hind limbs and the tail base.  Spatial zone occupancy (wall zone,
center zone) was determined by the position of the body center.  All
discrete variables were smoothed using a 0.25-second median filter to
remove noise and enforce minimal event duration.

\subsubsection{Processing of calcium imaging data}
\label{sec:fof-calcium-processing}

Calcium imaging data were processed using BEARMIND
(\url{https://github.com/iabs-neuro/bearmind}), a custom analysis
pipeline built on the CaImAn framework
\citep{giovannucci2019caiman}.  Preprocessing included motion
correction via NoRMCorre \citep{Pnevmatikakis2017} followed by source
extraction using CNMF-E \citep{Zhou2018}.  All retained components
were visually inspected to exclude artifacts and duplicates.
Fluorescence traces were normalized as $dF/F$.  The miniscope was not
removed between the three sessions of an animal, so the same neurons
were analyzed in each session.

In total, the dataset comprises 39 sessions from 13 animals (3
sessions per animal), with a mean of 373 simultaneously recorded
neurons per session (range 26--695; 14{,}533 neuron-sessions total)
at a frame rate of approximately 30~Hz.

\subsubsection{Analysis of the hippocampal dataset}
\label{sec:fof-analysis}

For Figure~\ref{fig:realdata}C, labels were counted after the
disentanglement stage of INTENSE \citep{Pospelov2026intense}, which
removes labels explained by covarying behavioral variables.  Concordance of prevalence rankings was
measured by Kendall's $W$ (without tie correction) over the
$13 \times 11$ matrix of per-mouse mean prevalences (percentage of
selective neurons, after the disentanglement stage), with $p$ from
$\chi^2 = m(n-1)W$ with 10 degrees of freedom.

Cross-session label retention (Results) was computed per feature,
after the disentanglement stage, as the fraction of neurons selective in at least one of an animal's three
sessions that were selective in all three.
A neuron counted as selective to a feature in a session if the pair was
significant at $p < 0.001$; no threshold on MI was applied.  Chance
retention was estimated by permuting labels between days: within each
animal and session, the labels were reassigned at random among the
matched neurons, keeping their number (10{,}000 permutations; one-sided $p = (b + 1)/(N + 1)$, where $b$ is the number of permutations in which at least as many neurons retained the label as observed and $N$ is the number of permutations).  Of 743 run-selective neurons, 19 (2.6\%) retained the label in all three sessions (1.7 expected by chance, 11-fold; $p < 0.001$), and of 788 place-selective neurons, 5 (0.6\%) did (1.7 expected, 3-fold; $p = 0.025$).  In at least two of the three sessions the label was retained by 114 run-selective neurons (58 expected) and 99 place-selective neurons (60 expected; both $p < 0.001$).

For Figure~\ref{fig:realdata}A we took the neurons of the example
session with a significant position, wall-zone or center-zone pair
(before the disentanglement stage), scaled each trace to $[0, 1]$, kept
every fifth frame and embedded the time points with UMAP (two
dimensions, 100 neighbors, minimum distance 0.9, Euclidean metric, no
fixed seed).  Procrustes disparity is the residual sum of squares after
both point sets are centered, scaled to unit norm and optimally rotated
or reflected.  Significance was assessed by PROTEST: position samples
were randomly reassigned to time points 999 times and
$p = (k+1)/1{,}000$, with $k$ the number of permutations with disparity
not exceeding the observed one; the test treats time points as
exchangeable.

\textbf{Scale-bridging control analysis.}  The filtering analysis of
Figure~\ref{fig:scalebridging}A was repeated on all sessions with at
least 10~spatially selective neurons (29 of 39~sessions from 11~mice),
with the same parameters as for the example session of
Figure~\ref{fig:scalebridging}A: downsampling $5\times$ and Isomap with
$k = 40$, two dimensions and a $k$-nearest-neighbor graph symmetrized by
union.  For each session we embedded all
neurons, the filtered set, 10~random subsets of neurons of the same
size as the filtered set, and three circular-shift controls each for
the all-neuron and filtered sets.  Calcium traces were min-max scaled
per neuron and decimated by taking every fifth frame (no averaging), and
random subsets were drawn without replacement from all recorded
neurons.  Held-out position $R^2$ was
computed by fitting the linear regression on four of five contiguous
time blocks and predicting the fifth, pooling residuals across folds;
contiguous blocks were used because neighboring frames are strongly
autocorrelated.  The embedding is unsupervised and was computed once on
all time points; only the linear read-out was cross-validated, with
$R^2 = 1 - \sum(\mathbf{p}-\hat{\mathbf{p}})^2 /
\sum(\mathbf{p}-\bar{\mathbf{p}}_{\mathrm{train}})^2$, so values below
zero are possible.  The spatially selective neurons forming the
filtered set were identified by INTENSE on the whole session.  Session values were
compared with paired Wilcoxon signed-rank tests, and per-mouse means
with Wilcoxon and sign tests; means are given with 95\% Student-$t$ confidence
intervals across sessions; all tests were two-sided.  Repeated runs of
the analysis are not bit-identical; among other sources, the
$k$-nearest-neighbor graph is built with an approximate search
(PyNNDescent) without a fixed seed.

Reverse analysis asked which recorded neurons carry information about
the population embedding.  In each session, the two coordinates of the
filtered embedding were linearly interpolated back to the original frame
rate and treated as two continuous features.  Every recorded neuron,
including those used to build the embedding, was tested against each
coordinate with the same two-stage test as for behavioral features
(GCMI, delays optimized within $\pm 2$~s, 100 and then 10{,}000 shifts),
and was called embedding-informative if it was significant for at least
one coordinate at $p < 0.001$.
For each session we report the fraction of embedding-informative
neurons that INTENSE classified as non-selective to all behavioral
features and compare it with the fraction of non-selective neurons in
the whole population.

The functional network of the example session
(Figure~\ref{fig:scalebridging}C) was built from calcium traces
downsampled five-fold.  All 241{,}165 neuron pairs were tested with the
two-stage circular-shift MI test (GCMI, zero lag, 100 and 1{,}000
shifts); a pair was connected by an unweighted edge if it passed both
stages ($p < 0.001$ at the validation stage), without multiple-comparison
correction.  About 241 edges are expected by chance at this threshold.
The shuffled control network was built in the same way from one
realization of independently shifted traces.  Spectra are the
eigenvalues of $L = I - D^{-1/2} A D^{-1/2}$ of the unweighted giant
component.  The primary selectivity type of a neuron is the group
(place; head or body direction; speed; run or walk; rest or freezing;
rearing; wall or center zone) of its significant feature with the
highest MI, with significance taken before the disentanglement stage.

Assortativity of the functional network by selectivity type was
computed as the attribute assortativity coefficient of the giant
component, with each neuron labelled by its primary selectivity type
and non-selective neurons forming one class; significance was assessed
by permuting the labels over the nodes 10{,}000 times with the graph
fixed.  To control for differences in the number of connections, the
permutation was repeated with labels exchanged only among nodes within
the same degree quantile stratum (six strata; four for the
selective-only subgraph).  The one-sided $p$-value was
$(b + 1)/(10{,}000 + 1)$, where $b$ is the number of permutations with
assortativity at least as large as the observed one.  The selective-only
analysis used the subgraph induced by the 232 selective neurons of the
giant component.

\subsection{Implementation and availability}

DRIADA is an open-source Python package released under the MIT License
and installable via \texttt{pip install driada}; it supports Python 3.9
through 3.13.  The analyses in this paper were run with version 1.2.1; release 1.2.2
changes only the scripts of this paper \citep{driada2026software}.
The source code is organized into six top-level modules
(\texttt{intense}, \texttt{dim\_reduction}, \texttt{network},
\texttt{rsa}, \texttt{information}, \texttt{recurrence}) plus
supporting modules for experiment management, synthetic data
generation, dimensionality estimation, and visualization.  An
experiment can be exported to an NWB (Neurodata Without Borders;
\citealp{Rubel2022nwb}) file and loaded back from it; the loader reads
NWB files in the layout that DRIADA itself writes and does not parse
NWB files produced by other tools.  The code is covered by automated
tests.  Documentation, including an API reference, tutorial notebooks
that run in Google Colab, and example scripts for each module, is
available at \url{https://driada.readthedocs.io}.

\subsection{Use of artificial intelligence tools}

AI coding assistants (Claude, Anthropic) were used to search for errors
in the software and to test it, and for language editing of the
manuscript.  The authors reviewed all outputs of
these tools and take full responsibility for the content of this
article.

\subsection{Materials availability}

This study did not generate new biological reagents; the software and
data generated in this study are available as described under Code
availability and Data availability.

\section{Funding}

This work was supported by the
Non-Commercial Foundation for Support of Science and Education
``INTELLECT.'' N.P. acknowledges support of IDEAS research center.
The funders had no role in study design, data collection and analysis,
interpretation, or the decision to submit the work for publication.

\section{Author contributions}

\textbf{Nikita Pospelov:} Conceptualization, Software, Methodology,
Validation, Formal analysis, Data curation, Writing -- original draft,
Writing -- review \& editing, Visualization, Project administration.
\textbf{Viktor Plusnin:} Software, Validation, Data curation, Writing
-- review \& editing. \textbf{Olga Rogozhnikova:} Software, Formal
analysis, Writing -- review \& editing. \textbf{Anna Ivanova:}
Investigation, Data curation, Writing -- review \& editing.
\textbf{Vladimir Sotskov:} Methodology, Investigation, Resources,
Writing -- review \& editing. \textbf{Margarita
Orobets:} Investigation, Data curation. \textbf{Ksenia Toropova:}
Methodology, Validation, Investigation, Data curation,  Writing --
review \& editing. \textbf{Olga Ivashkina:} Methodology, Validation, Investigation,
Data curation, Resources, Writing -- review \& editing, Visualization. \textbf{Vladik Avetisov:}
Formal analysis, Writing -- review \& editing. \textbf{Konstantin
Anokhin:} Supervision, Resources, Funding acquisition, Writing --
review \& editing.

\section{Competing interests}

The authors declare that no competing interests exist.

\section{Data availability}

The data analysed in this study, the fluorescence traces of all retained
neurons and the synchronized behavioral variables of all 39 sessions from
13 mice, are publicly available in the DANDI Archive as Dandiset 002030,
version 0.261016.1321
(\url{https://doi.org/10.48324/dandi.002030/0.261016.1321})
\citep{dandi002030}.

\section{Code availability}

\begin{sloppypar}
DRIADA (Dimensionality Reduction for Integrated Activity Data Analysis)
is available at \url{https://github.com/iabs-neuro/driada} under the MIT
License, with documentation at \url{https://driada.readthedocs.io}.
The analyses in this paper were run with version 1.2.1.  Release v1.2.2
\citep{driada2026software}, with the same library code, contains the
scripts that reproduce the figures, in the folder
\texttt{publications/elife}:
\url{https://github.com/iabs-neuro/driada/releases/tag/v1.2.2}.
\end{sloppypar}

The BEARMIND pipeline for miniscope video analysis is available at
\url{https://github.com/iabs-neuro/bearmind}.

The SPHYNX pipeline for automated behavioral analysis \citep{Plusnin1}
is available at
\url{https://github.com/iabs-neuro/sphynx}.

\section{Ethics}

All procedures were approved by the Commission of Bioethics of Lomonosov
Moscow State University (Application \textnumero{} 159-g, approved
during Bioethics Commission meeting \textnumero{} 154-d-r held on
17.08.2023), and followed the Russian Federation Order N~267~MZ and the
NIH Guide for the Care and Use of Laboratory Animals.

\bibliography{references_v10.1}

\end{document}